\documentclass[aps,twocolumn,nofootinbib]{revtex4}

\usepackage{graphicx}
\usepackage{dcolumn}
\usepackage{bm}

\usepackage{hyperref}
\hypersetup{
    colorlinks=true,
    linkcolor=blue,
    citecolor=blue,
    urlcolor=blue
}

\usepackage{amssymb,amsfonts}
\usepackage{epsfig}
\usepackage{epstopdf}
\usepackage[all]{xy}
\usepackage{amsthm}
\usepackage{url}
\usepackage{dsfont}
\usepackage{slashed}
\usepackage{mathrsfs}

\usepackage{bbm}
\usepackage{tikz}
\usepackage{physics}
\usepackage{enumitem}
\usepackage{soul}
\usepackage{float}

\newlength{\onecolwidth}
\AtBeginDocument{\setlength{\onecolwidth}{\dimexpr(\textwidth-\columnsep)/2\relax}}

\newcommand{\be}{\begin{align}}
\newcommand{\ee}{\end{align}}
\newcommand{\bea}{\begin{eqnarray}}

\newcommand{\eea}{\end{eqnarray}}

\def\inbar{\,\vrule height1.5ex width.4pt depth0pt}
\def\IR{\relax{\rm I\kern-.18em R}}
\def\IC{\relax\hbox{$\inbar\kern-.3em{\rm C}$}}

\begin{document}

\title{Infrared Spectral Gap in a Gluonic Dark Sector and the Galactic Acceleration Scale}

\author{Gilles Cohen-Tannoudji$^{1}$\footnote{gilles.cohentannoudji@gmail.com}} 

\author{Jean-Pierre Gazeau$^{2,3}$\footnote{gazeau@apc.in2p3.fr and j.gazeau@uwb.edu.pl}}

\author{Hamed Pejhan$^{4}$\footnote{pejhan@math.bas.bg}}

\author{Jean-Pierre Treuil$^{5}$\footnote{treuil.jean-pierre@orange.fr}}

\affiliation{$^1$Laboratoire de Recherche sur les Sciences de la Mati\`ere (LARSIM),  
CEA, Universit\'{e} Paris-Saclay, F-91190 Saint-Aubin, France}

\affiliation{$^2$Universit\'e Paris Cit\'{e}, CNRS, Astroparticule et Cosmologie, F-75013 Paris, France}

\affiliation{$^3$Faculty of Mathematics, University of Bia\l ystok, 15-245 Bia\l ystok, Poland}

\affiliation{$^4$Institute of Mathematics and Informatics, Bulgarian Academy of Sciences, Acad. G. Bonchev Str. Bl. 8, 1113, Sofia, Bulgaria}

\affiliation{$^5$Independent Researcher; formerly Research Engineer, Institut de Recherche pour le D\'{e}veloppement (IRD), France}

\date{\today}

\begin{abstract}
    {We further develop a trace-anomaly-motivated gluonic scenario in which cold dark matter (CDM) is modeled as a long-lived color-singlet Bose-Einstein condensate seeded at the QCD confinement transition. Specifically, guided by the near-universality of the galactic acceleration scale $g^{}_\dagger \simeq (1\text{--}2)\times10^{-10}\,\mathrm{m\,s^{-2}}$ inferred from the radial acceleration relation (RAR), we hypothesize that this relic gluonic condensate organizes, at galactic distances, into a spectrally rigid lowest-weight structure characterized by a protected infrared gap. Within an effective representation-theoretic framework, Lorentz covariance together with positive-energy lowest-weight unitarity naturally favors the Anti-de Sitter (AdS) algebra $\mathfrak{so}(2,3)$ as the minimal algebraic structure supporting such a discrete tower of states. The resulting condensate produces, by construction, a cored halo profile with finite total mass $M_h$. Normalizing this profile by the observed approximate universality of the central DM surface density, $\Sigma_0\simeq141\,M_\odot\,\mathrm{pc}^{-2}$, yields a universal characteristic acceleration $g^{}_\star ={G\, M_h}/{r_{\rm c}^2} \simeq \pi^2 G\,\Sigma_0 \simeq 1.9\times10^{-10}\,\mathrm{m\,s^{-2}}$, independent of the particular halo scale; since $\Sigma_0$ fixes the final expression, the dependence on the collective correlation length $r_{\rm c}$ drops out. Notably, its predicted value lies well within the empirical range of the acceleration scale inferred from the RAR. We also provide an illustrative comparison with representative rotation curves from the Spitzer Photometry and Accurate Rotation Curves (SPARC) database, indicating phenomenological compatibility of the finite-mass profile with realistic baryonic decompositions. This comparison is not a precision fit to the full SPARC sample, and the agreement with the RAR scale should not be interpreted as a first-principles derivation of the RAR itself. The framework links the QCD trace anomaly, gluonic dark matter, finite-mass cored halos, and the galactic acceleration scale within standard Newtonian gravity, without invoking modified gravity or fuzzy-DM quantum pressure.}
\end{abstract}

\maketitle

\setcounter{equation}{0} 
\section{Introduction}

The nature of cold dark matter (CDM) remains one of the central open problems in cosmology. In Ref. \cite{cohentan2021bec}, two of us developed a gluonic scenario in which CDM is modeled as a long-lived color-singlet Bose-Einstein condensate seeded at the QCD confinement transition. The microscopic input is the QCD trace anomaly \cite{Adler1982}, which plays two complementary roles.

\paragraph{First — dark/visible ratio.} Classical massless QCD is scale invariant, but this symmetry is broken by renormalization through the trace anomaly \cite{Adler1982}: 
\begin{align*}
    T^\mu{}_\mu = \frac{\beta(g_s)}{2g_s}\, G^a_{\mu\nu} G_a^{\mu\nu} + \cdots \,,
\end{align*}
with the one-loop beta function:
\begin{align*}
    \beta(g_s) = -\frac{g_s^3}{48\pi^2} \left(11N_c-2N_f\right) + \cdots \,.
\end{align*}
The coefficient contains two conceptually distinct contributions: a gluonic antiscreening term proportional to $11N_c$ and a quark screening term proportional to $2N_f$. The same balance is responsible for asymptotic freedom and underlies the running of the QCD coupling.

We posit that the gluonic component of the QCD trace anomaly provides the seed of a primordial confined dark sector, while the fermionic component is tied to the later baryonic visible sector. Accordingly, we use the relative anomaly weights as a minimal dimensionless guide to the dark-to-visible partition. This gives the reference value:
\begin{align*}
    \mathcal{R}_{\rm QCD} = \left. \frac{11N_c}{2N_f} \right|_{N_c=N_f=3} = \frac{11}{2} = 5.5 \,.
\end{align*}
This is close to the observed cosmological ratio:
\begin{align*}
    \mathcal{R}_{\rm obs} = \frac{\Omega_{\mathrm{DM}}}{\Omega_b} \simeq 5.36 \,,
\end{align*}
inferred from Planck cosmological parameters \cite{Planck2018}. We treat this numerical proximity not as a derivation, but as a phenomenological anomaly-sector matching ansatz. It identifies the QCD trace anomaly as a plausible microscopic organizing principle for the observed dark-to-baryonic abundance ratio, while its full cosmological justification would require a treatment of the non-equilibrium QCD transition, condensate survival, entropy production, baryogenesis, hadronization, and sector dilution.

\paragraph{Second — dynamical gluonic condensation.} Through dimensional transmutation, the anomaly breaks the classical scale invariance of Yang-Mills theory, for which $T^\mu{}_\mu=0$, and introduces a quantum theory characterized by an intrinsic confinement scale, $\Lambda_{\rm QCD}$. In non-Abelian gauge theories, dynamical scale generation naturally supports the formation of non-perturbative condensates with dimensions set by the generated scale. By analogy with the superconducting gap and its associated Cooper-pair condensate, a strongly coupled gauge theory may support not only fermionic condensates scaling as the third power of the dynamical scale, but also gluonic condensates scaling as its fourth power \cite{Adler1982}. This suggests that a residual color-singlet gluonic component may survive as a dynamically distinct infrared sector.

The present work addresses the collective level of this scenario. Rather than focusing on the microscopic origin of the gluonic component itself, we study its possible macroscopic infrared organization on galactic scales. A key phenomenological motivation is provided by the radial acceleration relation (RAR), whose tight regularity suggests that galactic dynamics may be controlled by an unexpectedly stable acceleration scale.

The RAR reveals a remarkably tight correlation between the observed acceleration in rotationally supported galaxies and that inferred from baryons, governed by a near-universal acceleration scale $g^{}_\dagger\simeq (1\text{--}2)\times10^{-10}\,\mathrm{m\,s^{-2}}$ with very small intrinsic scatter across diverse systems \cite{McGaugh2016, Lelli2017, lelli2016SPARC}. Although $\Lambda$CDM simulations incorporating baryonic feedback can reproduce similar RAR-like trends \cite{Ludlow2017, Keller2017, Navarro2017}, within the standard framework such regularity emerges statistically through baryon-halo coupling and galaxy assembly histories \cite{Navarro2017, Desmond2017, FamaeyMcGaugh2012}. If the stability of $g^{}_\dagger$ is confirmed by future analyses, notwithstanding recent studies questioning its statistical significance \cite{Julio2025}, it may instead indicate an intrinsic large-scale property of the dark sector.

This possibility naturally raises the central question of the present work: whether the RAR acceleration scale is the macroscopic manifestation of an underlying infrared scale governing DM dynamics. Since $g^{}_\dagger$ appears at kiloparsec distances, its origin is naturally sought in collective large-scale behavior rather than in ultraviolet particle physics alone. If the dark sector were effectively infrared scale-free at galactic distances, it would contain no obvious dynamically selected macroscopic length or acceleration. Halo structure could then be expected to depend more strongly on environmental factors — nonlinear collapse, mergers, and baryonic feedback — potentially leading to larger system-to-system variability. The empirical robustness of a single acceleration scale, however, motivates the existence of an intrinsic infrared regulator constraining macroscopic structure.

In quantum field theory and many-body systems, long-distance behavior is controlled by the lowest part of the spectrum. A gapless infrared spectrum typically leads to algebraic decay of correlations and selects no preferred macroscopic length scale. By contrast, a spectrum bounded from below by a non-zero gap produces exponential decay and dynamically introduces a finite correlation length \cite{Hastings2006}. The presence of such a gap, therefore, provides a natural mechanism for limiting large-distance coherence and selecting a characteristic structural scale.

However, a gap alone does not fully explain why this scale should remain rigid under nonlinear evolution and environmental variation. If the spectrum above threshold is continuous, long-distance behavior may still depend on near-threshold states and on the detailed spectral density, leaving room for system-dependent effects. A lowest-weight discrete structure is more restrictive; its ground state is isolated and spectrally protected, so macroscopic coherence is governed by a dominant infrared scale intrinsic to the dark sector. Although discreteness is not logically required, it provides a natural framework for understanding both the apparent universality and the small intrinsic scatter of the galactic acceleration scale.

In this view, self-gravitating equilibria inherit, rather than solely generate through environmental evolution, a characteristic structural scale, and the RAR is interpreted as the large-distance gravitational manifestation of this intrinsic infrared organization.

In the gluonic scenario adopted here, the QCD trace anomaly supplies a dimensionful microscopic infrared scale for the gluonic sector. However, neither the anomaly nor the associated condensate picture determines a macroscopic galactic correlation length or fixes the structure of the long-distance excitation spectrum. A theory with an intrinsic scale may still support gapless modes or continuous spectra in some channels. Thus, the anomaly provides a necessary microscopic seed for infrared organization, but not a sufficient mechanism for discreteness, boundedness from below, or spectral rigidity.

We therefore make the additional hypothesis that, at late cosmological times, the relevant color-singlet gluonic sector is organized by a lowest-weight unitary representation of $\mathfrak{so}(2,3)$. Confinement already motivates isolated color-singlet bound states; the further assumption is that, at sufficiently large scales, these states assemble into a Lorentz-covariant positive-energy structure. In four spacetime dimensions, a minimal and well-studied non-compact Lie algebra with this property is $\mathfrak{so}(2,3)$ \cite{fronsdal74,fronsdal75,enayati25, dSBook}. Its maximal compact subalgebra $\mathfrak{so}(2)\oplus\mathfrak{so}(3)$ contains a compact $\mathfrak{so}(2)$ generator with spectrum bounded from below, which plays the role of an effective Hamiltonian. The resulting lowest-weight spectrum is discrete, anchored at $E_0>0$, and characterized schematically by a spacing scale $\varkappa$, $E_k=E_0+k\hbar c\,\varkappa$, $k=0,1,2,\cdots$. This same scale defines an infrared correlation length $r_{\rm c}\sim{\varkappa}^{-1}$, with $\varkappa$ acting as an effective infrared mass parameter that controls the exponential falloff of color-singlet correlators \cite{Hastings2006}. The gap is protected within the lowest-weight unitary class, in the sense that it cannot be continuously removed without leaving this representation-theoretic organization. The appearance of $\mathfrak{so}(2,3)$ is therefore algebraic rather than geometric; no globally Anti-de Sitter (AdS) cosmological spacetime is assumed. The AdS algebra is used only to organize long-wavelength gluonic modes, and ``energy'' denotes the eigenvalue of the compact $\mathfrak{so}(2)$ generator, not the global energy of the Friedmann-Robertson-Walker background. In this way, the microscopic scale generated by dimensional transmutation is promoted to a spectrally rigid infrared length, whose gravitational imprint will be connected below to the observed galactic acceleration scale.

Given this lowest-weight infrared organization, a macroscopic occupation of the isolated ground state becomes natural. As in ordinary Bose-Einstein condensation, this does not require strong binding, but follows from Bose statistics together with the existence of a distinguished lowest mode. Here, this mode is intrinsic to the lowest-weight $\mathfrak{so}(2,3)$ structure rather than imposed by an external trap or finite-volume boundary condition. At the effective level, its occupation defines a self-gravitating condensate with this same correlation length.

The resulting condensate then yields, by construction, a cored halo profile with finite total mass $M_h$. The central point of the present paper is that this finite-mass profile converts the observed approximate universality of the central DM surface density \cite{KormendyFreeman2004, Donato2009, Gentile2009}\footnote{Although departures from this approximate universality have been reported in some systems, such as relaxed galaxy groups \cite{Gopika2021}, the existence of an approximately universal density-length product remains a robust phenomenological feature of cored halo descriptions.}:
\begin{align*}
    \Sigma_0 = \rho_c\, r_{\rm c} \,,
\end{align*}
where $\rho_c$ is the central DM density and $r_{\rm c}$ is the collective correlation length, into the characteristic acceleration:
\begin{align*}
    g^{}_\star = \frac{G M_h}{r_{\rm c}^2} \simeq \pi^2G\,\Sigma_0 \simeq1.9\times10^{-10} \,\mathrm{m\,s^{-2}} \,.
\end{align*}
This acceleration is fixed by the nearly universal density-length product $\Sigma_0$, is independent of halo size, and lies in the empirical range of the RAR acceleration scale. In this formulation, the scale is neither introduced as a fundamental constant nor fitted directly to rotation-curve data; rather, it emerges from combining a finite-mass cored halo profile with the observed near-universality of $\Sigma_0$. The agreement should therefore be viewed not as a first-principles derivation of the RAR itself, but as a possible dynamical explanation for the appearance of its characteristic acceleration scale within a coherent gluonic dark sector. A genuine first-principles derivation of the RAR would require, in addition, the condensate's linear response to baryonic perturbations, which lies beyond the scope of the present work. Such an analysis would provide an important test of the present interpretation and could clarify whether the observed scale $g^{}_\dagger$ can ultimately be determined from the microscopic parameters controlling the gluonic condensate, rather than inferred phenomenologically from galactic data.

The remainder of the paper develops the effective infrared construction, derives the associated halo profile, and connects its gravitational imprint to the galactic acceleration scale. We also present an illustrative comparison with representative SPARC rotation curves. This comparison is intended only as a phenomenological consistency check of the finite-mass profile against realistic baryonic decompositions; a full statistical analysis of the SPARC sample is left for future work.

\section{Gluonic Bose-Einstein Condensate in an Effective AdS Background} \label{GBEC}

\subsection{Effective AdS Spectrum and Scalar Casimir Equation} 

Restoring physical dimensions through the speed of light $c$ and Planck's constant $\hbar$, an effective AdS background with curvature radius $\varkappa^{-1}$ defines the frequency scale
\begin{align}
    \omega^{}_{\mathrm{AdS}} = \varkappa\, c \,.
\end{align}
The rest energy of a massive excitation with effective spin $s$ is then \cite{enayati25,dSBook}:
\begin{align} \label{restenAdS}
    E_{\mathrm{AdS}}^{\mathrm{rest}} = 
    \left[ m^2c^4 + \hbar^2\omega^2_{\mathrm{AdS}}\left(s - \frac{1}{2}\right)^2 \right]^{1/2} + \frac{3}{2}\hbar \omega^{}_{\mathrm{AdS}}\,,
\end{align}
and reduces to $E_{\rm Min}^{\rm rest}=mc^2$ in the flat limit $\varkappa\longrightarrow 0$.

As an effective working model, we consider non-interacting color-neutral scalar di-gluonic excitations with discrete spectrum:
\begin{align} \label{En}
    E_k = E_{\mathrm{AdS}}^{\mathrm{rest}} + k\hbar\omega^{}_{\mathrm{AdS}} 
    = \hbar\omega^{}_{\mathrm{AdS}}(k+\zeta) \,,
\end{align}
with $k=0,1,2,\cdots$, and:
\begin{align}\label{zeta}
    \zeta = \frac{E_{\mathrm{AdS}}^{\mathrm{rest}}}
    {\hbar\omega^{}_{\mathrm{AdS}}}\,.
\end{align}
The level degeneracy is:
\begin{align}\label{dege}
    G_k=\frac{(k+1)(k+2)}{2}\,.
\end{align}

The eigenstates relevant to the condensate model are scalar solutions of the fixed-Casimir equation \cite{fronsdal74, fronsdal75, enayati25, dSBook}:
\begin{align} \label{casAdS}
    \left(\mathrm{Q}_{\mathrm{AdS}}^{(1)}-\zeta(\zeta-3)\right) \varphi(x) = 0 \,,
\end{align}
where $\mathrm{Q}_{\mathrm{AdS}}^{(1)}$ is the quadratic Casimir of $\mathfrak{so}(2,3)$. The parameter $\zeta$ fixes the corresponding scalar lowest-weight unitary irreducible representation. Higher-spin color-singlet modes are neglected in the present effective description, under the assumption that the leading infrared organization is governed by the lowest-weight scalar sector.

In global static coordinates $x=(t,\mathbf{x})$, with $\mathbf{x}=(r,\theta,\phi)$ and $t\in\mathbb R$ on the universal covering of AdS, the metric takes the form:
\begin{align} \label{globalmetric}
    \mathrm{d}s^2 = \left(1+(\varkappa r)^2\right)c^2\mathrm{d}t^2 - \frac{\mathrm{d}r^2}{1+(\varkappa r)^2} - r^2\mathrm{d}\Omega_2^2 \,,
\end{align}
where $\mathrm{d}\Omega_2^2$ denotes the metric on the unit two-sphere. For scalar fields, the Casimir equation \eqref{casAdS} is equivalently the Klein-Gordon-type equation:
\begin{align} \label{KGAdS}
    \left(\Box_{\mathrm{AdS}} + \varkappa^2\zeta(\zeta-3)\right)\varphi(x) = 0 \,.
\end{align}

\subsection{AdS Modes and Their Normalization} 

\subsubsection{General Expression}
   
The normal-mode solutions of Eq. \eqref{KGAdS} were obtained by Fronsdal in Ref. \cite{fronsdal74}. On the universal covering of AdS, where the global time coordinate is unwrapped, $t\in\mathbb R$, these modes may be considered for the continuous range $\zeta>3/2$ and are given by:
\begin{align} \label{waveads}
    \varphi^{}_{nlm}(x) =&\, C_{nl}\,Y_{lm}\! \left(\theta,\phi\right)\, \mathrm{e}^{-\mathrm{i} (\zeta + 2n +l)\omega^{}_{\mathrm{AdS}} t} \nonumber\\
    &\times (\varkappa r)^{2n+l}\, \left(1 + (\varkappa r)^2\right)^{-\frac{1}{2}(\zeta +2n + l)} \nonumber\\
    &\times {}_{2}F^{}_1\!\left(-n,-n-l -\frac12;\, \zeta -\frac12;\, -\frac{1}{(\varkappa r)^2} \right)\,,
\end{align}
with:
\begin{align} \label{Ckl}
    C_{nl} =&\, \sqrt{\frac{c}{\pi}}\frac{\varkappa^2} {\Gamma\left(\zeta -\frac{1}{2}\right)}\nonumber\\
    &\times \left[\frac{\left(\zeta-\frac{3}{2}\right)\Gamma(\zeta+n+l)\Gamma\left(\zeta + n - \frac{1}{2}\right)}{\Gamma\left(n +l +\frac{3}{2}\right)\Gamma(n+1)}\right]^{1/2} \,.
\end{align} 
The spherical harmonics $Y_{lm}$ are normalized on $\mathbb S^2$, and the quantum numbers satisfy $n,l\in\mathbb{N}=\left\{0,1,2, \cdots \right\}$ and $-l\leq m\leq l$. Each mode $\varphi^{}_{nlm}$ is a simultaneous eigenfunction of the AdS energy generator $L_{05}$, of the $\mathfrak{so}(3)$ Casimir $\mathbf{L}^2$, and of $L_{12}$. The corresponding eigenvalues are $\zeta+l+2n$, $l(l+1)$, and $m$, respectively. On AdS itself, the global time is periodic, $t\sim t+2\pi/\omega_{\mathrm{AdS}}$. Single-valuedness of the phase factor
$\exp[-\mathrm{i}(\zeta+2n+l)\omega_{\mathrm{AdS}}t]$ then requires $\zeta\in\mathbb Z$, giving the scalar positive-energy values $\zeta=2,3,\cdots$. Passing to the universal covering removes this periodicity condition and allows the broader range $\zeta>3/2$. The function $^{}_2F^{}_1(a,b;c;r)$ denotes the Gaussian hypergeometric function. We recall the following standard properties, which will be used below in analyzing the behavior of the lowest mode and the truncation of the corresponding radial profiles:
\begin{align}
    ^{}_2F^{}_1(a,\,b;\,c;\,r) &= \frac{\Gamma(c)}{\Gamma(a)\, \Gamma(b)} \;\sum_{p=0}^{\infty} \frac{\Gamma(a+p)\, \Gamma(b+p)}{\Gamma(c+p)} \; \frac{r^p}{p!} \nonumber\\
    &= 1 + \frac{a\,b}{c}\, \frac{r}{1!} + \frac{a(a+1)\, b(b+1)}{c(c+1)} \frac{r^2}{2!} + \cdots \,,
\end{align}
for $c\neq 0,-1,-2,\cdots$. This series possesses the following properties:
\begin{enumerate}
    \item{It converges in the open unit disk $|r| < 1$.}
    
    \item{Its behavior on the circle of convergence $|r| =1$ is given by:
    \begin{enumerate}
        \item{divergent for $\mbox{Re} (a+b-c) \geqslant 1$,}
        
        \item{absolutely convergent for $\mbox{Re} (a+b-c) < 0$,}
        
        \item{conditionally convergent for $0 \leqslant \mbox{Re} (a+b-c) < 1$, the point $r =1$ being excluded.}
     \end{enumerate}}
     
    \item{It reduces to a polynomial of degree $p$ in $r$, when $a$ or $b$ is a negative integer $-p$ ($p = 0,1,2,3, \cdots$).}
\end{enumerate}

The natural invariant measure in global AdS coordinates may be written as:
\begin{align} \label{muads}
    \mathrm{d}\mu(x) = \sqrt{-g}\,\mathrm{d}^4x = r^2\sin\theta\, \mathrm{d}t\, \mathrm{d}r\, \mathrm{d}\theta\, \mathrm{d}\phi \,.
\end{align}
With the normalization constant $C_{nl}$ chosen as in \eqref{Ckl}, the modes satisfy the standard Fronsdal orthonormality relation:
\begin{align} \label{basisads}
    \int_{\mathrm{AdS}} \overline{\varphi^{}_{nlm}(x)}\, \varphi^{}_{n^{\prime}l^{\prime}m^{\prime}}(x)\, \mathrm{d}\mu(x) =  \delta_{nn^{\prime}}\, \delta_{ll^{\prime}}\, \delta_{mm^{\prime}}\,.
\end{align}
The Fronsdal orthonormality relation \eqref{basisads} is naturally formulated on AdS with compact global time. On the universal covering, where $t\in\mathbb R$, the time direction becomes non-compact, so the full spacetime normalization no longer has the same finite-volume form. This distinction is harmless for the present purpose, since the condensate density depends only on the spatial profile $|\varphi^{}_{nlm}(t,\mathbf{x})|^2$, and the time-dependent phase drops out.

\subsubsection{Ground State and First Excited Angular Modes}

\paragraph{Spherically symmetric ground state.} The expression for the ground state explicitly reads as:
\begin{align} \label{ground}
    \varphi^{}_{000}(x) = \frac{\varkappa^2 \sqrt{c}}{2\pi} \left[B\left(\frac{3}{2},\zeta - \frac{3}{2}\right)\right]^{-1/2} \frac{\mathrm{e}^{-\mathrm{i} \zeta \omega^{}_{\mathrm{AdS}} t}}{\left(1+ (\varkappa r)^2\right)^{\frac{\zeta}{2}}} \,,
\end{align}
where $B(p,q)=\Gamma(p)\Gamma(q)/\Gamma(p+q)$ denotes the Euler (complete) beta function. Note that $Y_{00}=1/\sqrt{4\pi}$. 

\paragraph{First dipolar angular excitations.} The first angularly excited states are:
\begin{align} \label{01m}
    \varphi^{}_{01m}(x) =&\, \varkappa^{2}\,\sqrt{\frac{c}{\pi}}\,\left[B\! \left(\frac{5}{2},\,\zeta-\frac{3}{2}\right) \right]^{-1/2} \nonumber\\
    & \times Y_{1m}(\theta,\phi)\, \mathrm{e}^{-\mathrm{i} (\zeta + 1)\omega^{}_{\mathrm{AdS}} t}\, \frac{\varkappa r}{\left(1+ (\varkappa r)^2\right)^{\frac{1}{2}(\zeta + 1)}}\,,
\end{align}
where $m=-1,0,1$.

\paragraph{Higher radial excitations.} For $n\geq 1$, the hypergeometric function appearing in \eqref{waveads} truncates to a polynomial of degree $n$ in $1/(\varkappa r)^2$. Consequently, the radial dependence can be written as a finite sum of terms of the form:
\begin{align}\label{nGe1}
    \frac{(\varkappa r)^{2n+l-2p}}{\left(1+(\varkappa r)^2\right)^{\frac{1}{2}(\zeta+2n+l)}} \,,
\end{align}
with $p=0,\cdots,n$. 

Since our analysis focuses primarily on the ground state and the lowest excited levels, we do not pursue the general structure of these higher-order polynomials further here.

\subsection{The Low-$\zeta$ Sector, Boundary Behavior, and Conformal Scalar Structure}

\paragraph{Breakdown of the $L^2$ realization and its remedy.} It is important to emphasize that, with respect to the standard inner product \eqref{basisads}, the interval:
\begin{align}
    \tfrac12 \le \zeta \le \tfrac32
\end{align}
contains modes that fail to be square-integrable in the ordinary $L^2$ sense. This loss of square integrability does not signal any pathology of the  corresponding representations; rather, it indicates that the naive Hilbert-space realization is no longer adequate in this regime.

A consistent framework is nevertheless available. The singleton (Rac) case $\zeta=\tfrac12$, as well as the intermediate range $\tfrac12 < \zeta \le \tfrac32$, can be treated by replacing the standard $L^2$ norm with a Klein-Gordon-type inner product, suitably renormalized at the AdS boundary, and by defining the physical state space as a quotient by null states (see Ref. \cite{fronsdal75}). 

At the level of the radial integration, this change is reflected in a modification of the effective measure; in the range $\tfrac12 < \zeta \le \tfrac32$, the standard weight $x^2\,\mathrm{d}x$ is replaced by:
\begin{align}
    x^2(1+x^2)^{-1}\,\mathrm{d}x\,,
\end{align}
which compensates for the slower decay of the modes near the boundary. The endpoint $\zeta = \tfrac12$ (Rac) requires a separate and more delicate treatment. This distinction reflects the different boundary behavior of the modes and is fully consistent with the renormalized Klein-Gordon framework appropriate to the low-$\zeta$ sector.

\paragraph{AdS/CFT interpretation and double quantization.} A closely related manifestation of this structure appears in the AdS/CFT correspondence. For scalar modes in four-dimensional AdS, the near-boundary behavior involves two independent fall-offs:
\begin{align}
    \phi(r)\sim r^{-\Delta_-}\,, &\quad \Delta_-=\zeta\,, \nonumber\\
    \phi(r)\sim r^{-\Delta_+}\,, &\quad \Delta_+=3-\zeta\,.
\end{align}
In the window $\tfrac12<\zeta\le\tfrac32$, the slower branch is $r^{-\zeta}$, while the faster branch is $r^{-(3-\zeta)}$. 
This interval is exactly the regime in which both branches are admissible with respect to a suitably renormalized Klein-Gordon inner product.

The choice of Hilbert-space structure then selects a quantization; the standard $L^2$ construction selects the faster-decaying branch $\Delta_+=3-\zeta$ (standard quantization), while the renormalized framework allows one to include the slower-decaying branch $\Delta_-=\zeta$ consistently (alternate quantization). The endpoint $\zeta=\tfrac12$ gives $(\Delta_-,\Delta_+)=(\tfrac12,\tfrac52)$ and reproduces the Rac singleton.

\paragraph{Conformal scalar and the special role of $\zeta=2$.} A particularly important situation arises for the conformally coupled scalar in four-dimensional AdS. The complementary values $\zeta=2$ and $\zeta=1$ yield the same Casimir eigenvalue:
\begin{align}
    \zeta(\zeta-3)=-2 \,,
\end{align}
and therefore correspond to the same conformally invariant wave equation. They should thus be viewed as two realizations of a single underlying conformal structure associated with the reducible $\mathfrak{so}(2,3)$ module $D(2,0)\oplus D(1,0)$, whose components are naturally mixed under the action of the full conformal algebra.

Their distinction lies in their boundary behavior; the $\zeta=2$ modes correspond to the faster fall-off (standard quantization), whereas the $\zeta=1$ modes realize the slower branch (alternate quantization). In modern terms, this system lies in the Breitenl\"{o}hner-Freedman window and admits a one-parameter family of self-adjoint extensions, equivalently Robin boundary conditions, with the pure $\zeta=2$ and $\zeta=1$ realizations recovered as limiting cases.

\paragraph{Selection of the finite-mass sector.} For the halo construction developed below, we shall focus on the standard normalizable sector $\zeta>3/2$. In this range, the ground-state profile leads, when interpreted as a physical halo density, to a finite total mass. The finite-mass condition and its phenomenological consequences will be discussed explicitly in Sec. \ref{subsec:mass_profiles}. The value $\zeta=2$ then emerges as the conformal partner compatible with the finite-mass sector, while $\zeta=1$ remains the complementary conformal branch in the full boundary-condition problem.

\section{DM Mass Profiles, Rotation Curves, and Accelerations} \label{DMProf}

In this section, we examine the main properties of the corresponding DM halo structure, including the mass profile, circular velocity, and radial acceleration. For clarity, we first restrict attention to a purely spherically symmetric configuration dominated by the ground state \eqref{ground}. Within the AdS-Bose-Einstein condensate framework described above, the associated DM density profile is taken to be:
\begin{align}\label{density}
    \rho^{}_{\mathrm{DM}}(x)=\rho_c(1+x^2)^{-\zeta}\,, \quad x=\varkappa r=\frac{r}{r_{\rm c}} \,.
\end{align}
Here, $r_{\rm c}=\varkappa^{-1}$ is the collective correlation length of the condensate, while the phenomenological normalization is fixed by the observed surface-density scale $\Sigma_0=\rho_c r_{\rm c}$ \cite{KormendyFreeman2004, Donato2009, Gentile2009}.

It is important to emphasize, once again, that the effective AdS geometry is not interpreted as the physical spacetime geometry of the galactic halo. It is used as an auxiliary representation-theoretic structure organizing the infrared spectrum and normal modes of the dark sector. The resulting ground-state profile is then interpreted in ordinary physical space by identifying $x=\varkappa r=r/r_{\rm c}$, where $r$ is the usual weak-field radial coordinate. Accordingly, the enclosed mass, circular velocity, and radial acceleration are computed with the standard Newtonian measure $\mathrm{d}^3\mathbf{x}=4\pi r^2\,\mathrm{d}r$.

\subsection{Mass Profile Properties} \label{subsec:mass_profiles}

For fixed $(\varkappa,\zeta)$, integrating the density profile \eqref{density} over a sphere of radius $r$, with $x=\varkappa r=r/r_{\rm c}$, gives:
\begin{align} \label{eq:Mx}
    M_{\mathrm{DM}}(<x,\zeta) = 4\pi \rho^{}_{\texttt{c}} r_{\rm c}^3 F(x,\zeta) \,,
\end{align}
where the dimensionless mass function $F(x,\zeta)$ is defined by:
\begin{align} \label{eq:Fdef}
    F(x,\zeta) = \int_0^x f(u,\zeta) \,\mathrm{d}u \,, \quad f(u,\zeta) = u^2(1+u^2)^{-\zeta} \,.
\end{align}

The function $F(x,\zeta)$ can be expressed in closed form by introducing the variable:
\begin{align}
    t = \frac{u^2}{1+u^2} \,,
\end{align}
for which:
\begin{align}
    u^2 = \frac{t}{1-t}\,, \quad \mathrm{d}u = \frac{\mathrm{d}t}{2\,t^{1/2}(1-t)^{3/2}} \,.
\end{align}
Substituting into Eq. \eqref{eq:Fdef}, one finds:
\begin{align}\label{integral}
    F(x,\zeta) &=\int_0^x u^2(1+u^2)^{-\zeta}\,\mathrm{d}u \nonumber\\
    &=\frac12 \int_0^{\frac{x^2}{1+x^2}}
    t^{1/2}(1-t)^{\zeta-\frac52}\,\mathrm{d}t \,.
\end{align}
This is precisely the incomplete beta function:
\begin{align}
    B_z(a,b) = \int_0^z t^{a-1}(1-t)^{b-1}\,\mathrm{d}t \,,
\end{align}
with $a = 3/2$, $b=\zeta-3/2$, and $z=x^2/(1+x^2)$. Therefore:
\begin{align} \label{eq:Fbeta}
    F(x,\zeta) &= \frac12\,B_{\frac{x^2}{1+x^2}} \!\left(\frac32,\zeta-\frac32\right) \nonumber\\
    &= \frac13\!\left(\frac{x^2}{1+x^2}\right)^{3/2}
    {}_2F^{}_1\!\left(\frac32,\, \frac52-\zeta;\, \frac52;\, \frac{x^2}{1+x^2}\right)\,,
\end{align}
where in the second line we used the standard hypergeometric representation of the incomplete beta function:
\begin{align}
    B_z(a,b) = \frac{z^a}{a}\, {}_2F^{}_1(a,\, 1-b;\, a+1; \,z) \,.
\end{align}

For $\zeta>\frac32$, $F(x,\zeta)$ converges as $x\longrightarrow\infty$ to the finite limit:
\begin{align} \label{eq:Finf}
    F(\infty,\zeta) = \frac12\,B\!\left(\frac32,\zeta - \frac32\right) \,.
\end{align}
Consequently, the total mass of the AdS-gluonic-condensate halo is finite by construction and is given by:
\begin{align} \label{eq:Mh}
    M_h = M_{\mathrm{DM}}(<\infty, \zeta) = 4\pi\rho^{}_{\texttt{c}} r_{\rm c}^3\; \frac12\,B\!\left(\frac32,\zeta-\frac32\right) \,.
\end{align}

This is the precise sense in which the condition $\zeta>3/2$ selects the finite-mass sector used in the preceding mode construction. This property is not uniform, however, as $\zeta$ approaches the critical value $\zeta=3/2$. Indeed, in the near-critical regime $\zeta=3/2+\epsilon$, with $\epsilon>0$, the density profile approaches the familiar outer behavior $\rho(r)\sim r^{-3}$, while $F(\infty,\zeta)$ remains finite only for each fixed non-zero $\epsilon$. In the limit $\epsilon\longrightarrow0^+$, Eq. \eqref{eq:Finf} diverges, and the total mass becomes increasingly dominated by the outer tail of the profile.

The near-critical regime $\zeta\simeq3/2$ is therefore phenomenologically interesting, since it connects the condensate profile to the standard $r^{-3}$-type outer falloff. At the same time, it lies at the boundary of the finite-mass sector, where the contribution of the outer halo region to the total mass becomes increasingly significant. As will be shown below, this regime tends to describe more extended and diffuse halo configurations.

Away from this critical boundary, and in particular for the benchmark values $\zeta\geq2$ considered below, the finite-mass property is a central structural advantage of the present profile. Although the density distribution extends formally to infinite radius, the total halo mass remains finite. This contrasts with several commonly used phenomenological halo profiles (see, for instance, Refs. \cite{Begeman1991, Burkert1995, Navarro1997, Read2016, Li2020}), whose enclosed mass diverges logarithmically or linearly at large radii unless an explicit truncation radius is imposed.

Against this background, $\zeta=2$ provides a natural reference value: it is the finite-mass member of the conformal scalar pair $\zeta=1,2$, lies safely above the critical boundary, and retains analytic simplicity.

\paragraph{Regularized form.} It is convenient to express the mass profile in a normalized (regularized) form:
\begin{align} \label{eq:Mreg}
    \frac{M_{\mathrm{DM}}(<x,\zeta)}{M_h} = \frac{F(x,\zeta)}{F(\infty,\zeta)} = \frac{B_{\frac{x^2} {1+x^2}}\!\left(\frac32,\zeta-\frac32\right)}{B\!\left(\frac32,\zeta-\frac32\right)} \,.
\end{align}

\paragraph{Central and global behavior.} Near the halo center ($x\ll1$), using $F(x,\zeta)\sim x^3/3$, one recovers a constant-density core:
\begin{align} \label{eq:core}
    M_{\mathrm{DM}}(<x,\zeta)\sim \frac{4}{3}\pi\rho^{}_{\texttt{c}} r_{\rm c}^3\,x^3 \,.
\end{align}

\paragraph{Mass concentration and $\zeta$-dependence of the halo profile.} The function $f(x,\zeta)=x^2(1+x^2)^{-\zeta}$ reaches its maximum at:
\begin{align} \label{eq:xmax}
    x^{}_{M;\max} = (\zeta-1)^{-1/2} \,,
\end{align}
as follows from the condition $\mathrm{d}(\ln{f})/\mathrm{d}x=0$, equivalent to $f^\prime(x,\zeta)=0$ for $x>0$. Since $F^\prime(x,\zeta) = f(x,\zeta)$, this point coincides with the inflection point of $F(x,\zeta)$. The maximal value is:
\begin{align}\label{eq:fxmax}
    f_{\max}=\frac{(\zeta-1)^{\zeta-1}} {\zeta^\zeta} \,.
\end{align}
Recall that $\zeta>3/2$.

These properties show that, for fixed $\rho_c$ and $r_{\rm c}$, increasing $\zeta$ leads to a stronger central concentration. Indeed, for the density profile \eqref{density}, larger values of $\zeta$ produce a progressively steeper fall-off at large radii:
\begin{align}\label{eq:xmax'}
    (1+x^2)^{-\zeta}\sim x^{-2\zeta}\,, \quad \mbox{for} \quad x\gg1 \,.
\end{align}
Consequently, the outer halo contributes less to the mass integral, and a larger fraction of the total mass becomes confined within the inner region $r\lesssim r_{\rm c}$. Consistently, the total halo mass decreases asymptotically as:
\begin{align}\label{mh zeta>>1}
    M_h \sim \pi^{3/2}\rho_c r_{\rm c}^3\,\zeta^{-3/2}\,, \quad \mbox{for} \quad \zeta\gg1 \,.
\end{align}

\begin{figure*}[t]
    \centering
    \includegraphics[width=0.55\linewidth]{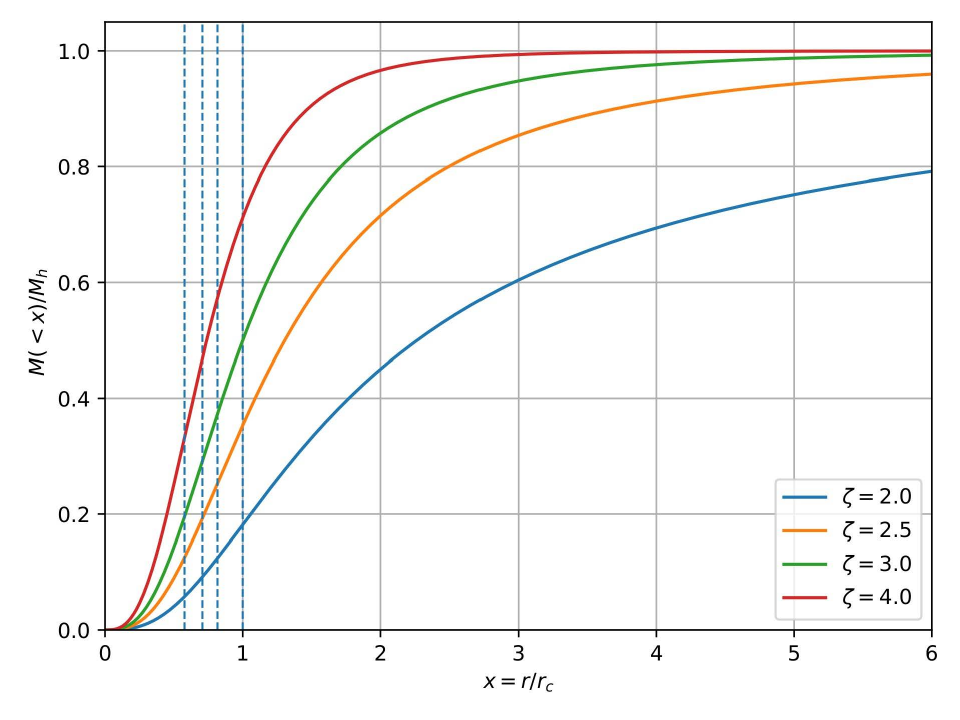}
    \caption{Normalized enclosed mass $M_{\mathrm{DM}}(<x,\zeta)/M_h$ as a function of $x=r/r_{\rm c}$ for $\zeta=2,\;5/2,\;3,\;4$. Vertical dashed lines indicate the inflection radii $x^{}_{M;\max} = (\zeta-1)^{-1/2}$. Larger values of $\zeta$ correspond to a stronger central concentration and a faster saturation of the total halo mass.}
    \label{fig:mass_profiles1}
\end{figure*}

\begin{figure*}[t]
    \centering
    \includegraphics[width=0.55\linewidth]{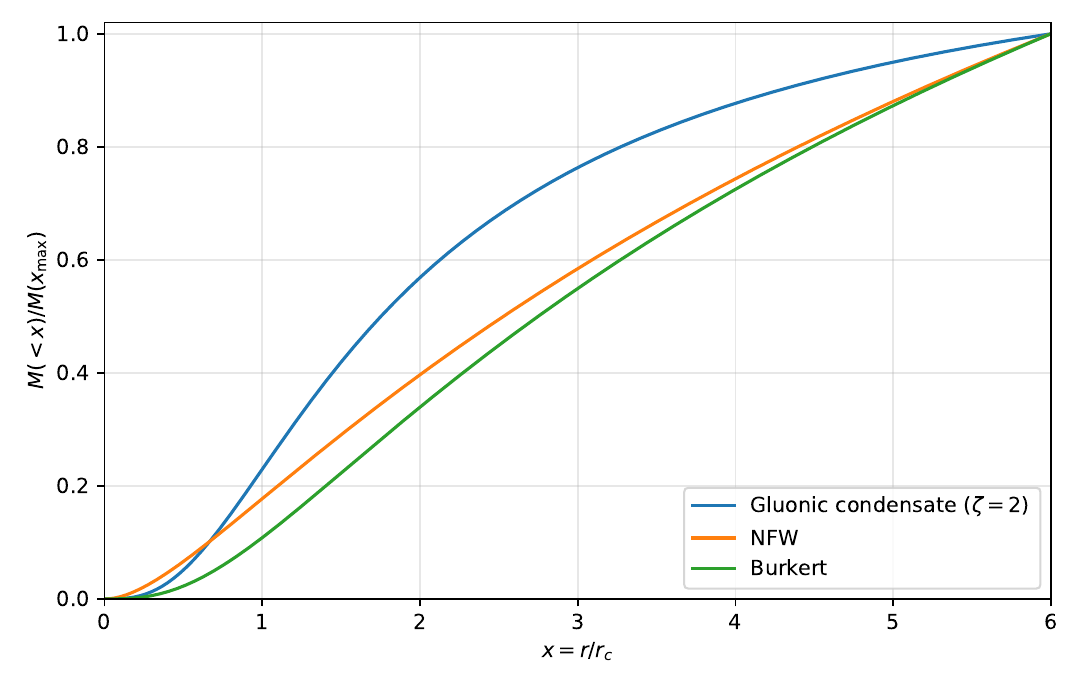}
    \caption{Comparison of normalized enclosed mass profiles. The AdS-gluonic-condensate halo (here $\zeta=2$) is compared with the Navarro-Frenk-White (NFW) and Burkert profiles, all normalized at the same outer radius $x^{}_{M;\max}=6$. $M(<x)=M_{\mathrm{DM}}(<x,\zeta)$ for the AdS-gluonic-condensate halo. Unlike phenomenological profiles, the present model naturally leads to a finite total mass without requiring an explicit cutoff.}
    \label{fig:mass_profiles2}
\end{figure*}

Figure \ref{fig:mass_profiles1} shows the normalized enclosed mass $M_{\mathrm{DM}}(<x,\zeta)/M_h$ as a function of $x=r/r_{\rm c}$ for $\zeta=2,\;5/2,\;3,\;4$. Figure \ref{fig:mass_profiles2} compares the normalized enclosed mass profile of the AdS-gluonic-condensate halo (for $\zeta=2$) with those of the Navarro-Frenk-White (NFW) and Burkert profiles.

A key structural distinction emerges, summarized in Table \ref{tab:halo_profiles}. In widely used phenomenological halo models, the enclosed mass diverges at large radii and a finite halo mass can only be defined by introducing an arbitrary truncation scale (typically $R_{200}$). In contrast, the AdS-gluonic-condensate profile yields a finite total mass for $\zeta>\tfrac{3}{2}$ despite its formally infinite spatial support. The halo mass, therefore, is determined intrinsically by the profile itself, rather than from an externally imposed cutoff.

\begin{table*}[t]
    \centering
    \renewcommand{\arraystretch}{1.3}
    \begin{tabular}{|l|c|c|c|c|c|}
    \hline
    \textbf{Profile}
    & $\rho(r\!\longrightarrow\!0)$
    & $\rho(r\!\longrightarrow\!\infty)$
    & Core / Cusp
    & $M(<r\!\longrightarrow\!\infty)$
    & Reference \\ \hline

    Pseudo-isothermal
    & finite
    & $\sim r^{-2}$
    & Core
    & $\propto r$
    & Begeman et al.\ (1991) \\ \hline

    Burkert
    & finite
    & $\sim r^{-3}$
    & Core
    & $\sim \ln r$
    & Burkert (1995) \\ \hline

    NFW
    & $\sim r^{-1}$
    & $\sim r^{-3}$
    & Cusp
    & $\sim \ln r$
    & Navarro et al.\ (1997) \\ \hline

    coreNFW
    & finite
    & $\sim r^{-3}$
    & Core
    & $\sim \ln r$
    & Read et al.\ (2016) \\ \hline

    This work (Gluonic condensate)
    & finite
    & $\sim r^{-2\zeta}$
    & Core
    & \textbf{finite} for $\zeta>\tfrac32$
    & Present work \\ \hline

    \end{tabular}
    \caption{Comparison of commonly used DM halo density profiles. Here, $\rho(r)$ denotes the standard three-dimensional mass density, so that the enclosed mass is $M(<r) = 4\pi\int_0^r \rho(r^\prime)\, {r^\prime}^2\, \mathrm{d}r^\prime$. The asymptotic behaviors quoted in the table refer to this volumetric density. Profiles such as pseudo-isothermal, Burkert, and NFW either lead to a linearly or logarithmically divergent total mass as $r\longrightarrow\infty$, unless an explicit outer cutoff radius is introduced. By contrast, the AdS gluonic-condensate profile considered in this work, $\rho(r)\sim r^{-2\zeta}$, yields a finite total mass for $\zeta>\tfrac{3}{2}$ without requiring any \emph{ad hoc} truncation, thereby providing a self-consistent and intrinsically regular halo model.}
    \label{tab:halo_profiles}
\end{table*}

\subsection{Rotation Curves} \label{subsec:rotation_curves}

The circular velocity induced by the DM halo in its ground state follows from the usual relation for a spherically symmetric mass distribution:
\begin{align} \label{eq:vdef}
    v^2(x,\zeta) = \frac{G M_{\mathrm{DM}}(<x,\zeta)}{r_{\rm c}\,x} = 4\pi G\rho^{}_{\texttt{c}} r_{\rm c}^2\,\frac{F(x,\zeta)}{x} \,,
\end{align}
where $x=r/r_{\rm c}$ and $F(x,\zeta)$ is the dimensionless enclosed-mass function introduced in Sec. \ref{subsec:mass_profiles}. It is convenient to write the velocity profile explicitly as:
\begin{align} \label{eq:vexplicit}
    v(x,\zeta) = \sqrt{4\pi G\rho^{}_{\texttt{c}} r_{\rm c}^2\,\frac{F(x,\zeta)}{x}} \,.
\end{align}
Differentiating with respect to the dimensionless radius $x$ yields:
\begin{align} \label{eq:vdx}
    \frac{\mathrm{d}v(x,\zeta)}{\mathrm{d}x} = \frac{4\pi G\rho^{}_{\texttt{c}} r_{\rm c}^2}{2\,v(x,\zeta)} \left( - \frac{F(x,\zeta)}{x^2} + \frac{f(x,\zeta)}{x} \right) \,,
\end{align}
where $f(x,\zeta)=x^2(1+x^2)^{-\zeta}$ is the integrand appearing in the definition of $F(x,\zeta)$.

\paragraph{Central behavior.} Near the halo center ($x\ll1$), using $F(x,\zeta)\sim x^3/3$, one finds:
\begin{align} \label{eq:vcore}
    v(x,\zeta) \sim \sqrt{\frac{4\pi G}{3}\rho^{}_{\texttt{c}}}\; r_{\rm c}\,x \,.
\end{align}
Thus, the velocity grows linearly with radius, $v\propto r = r_{\rm c}\,x$, which is the characteristic behavior of a solid-body rotation curve.

\paragraph{Asymptotic behavior.} For $x\gg 1$ and $\zeta>\tfrac32$, the function $F(x,\zeta)$ approaches its asymptotic value $F(\infty,\zeta)$ \eqref{eq:Finf}. The circular velocity then becomes (see Eq. \eqref{eq:Mh}):
\begin{align} \label{eq:vasymp}
    v(x,\zeta) \sim \sqrt{\frac{G M_h(\zeta)}{r_{\rm c}\,x}} \,,
\end{align}
which corresponds to a slowly declining Keplerian behavior $v\propto r^{-1/2}$. For fixed $(\rho^{}_{\texttt{c}},r_{\rm c})$, increasing $\zeta$ leads to a smaller total halo mass $M_h(\zeta)$ (see Eq. \eqref{mh zeta>>1}), and consequently to systematically lower asymptotic velocities.

\begin{figure*}[t]
    \centering
    \includegraphics[width=0.49\linewidth]{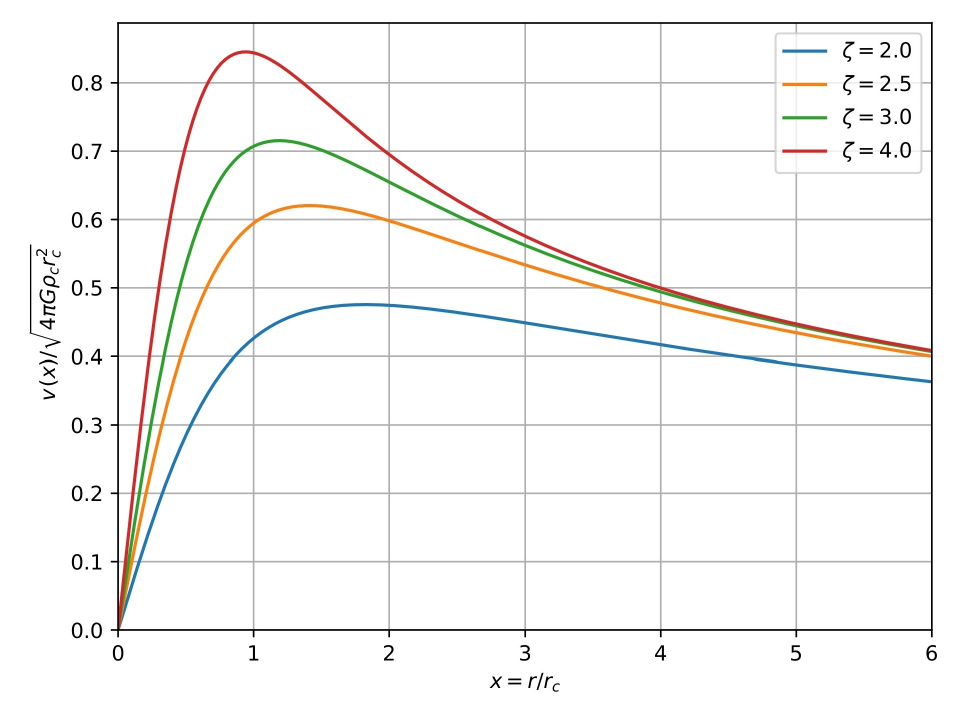}
    \includegraphics[width=0.49\linewidth]{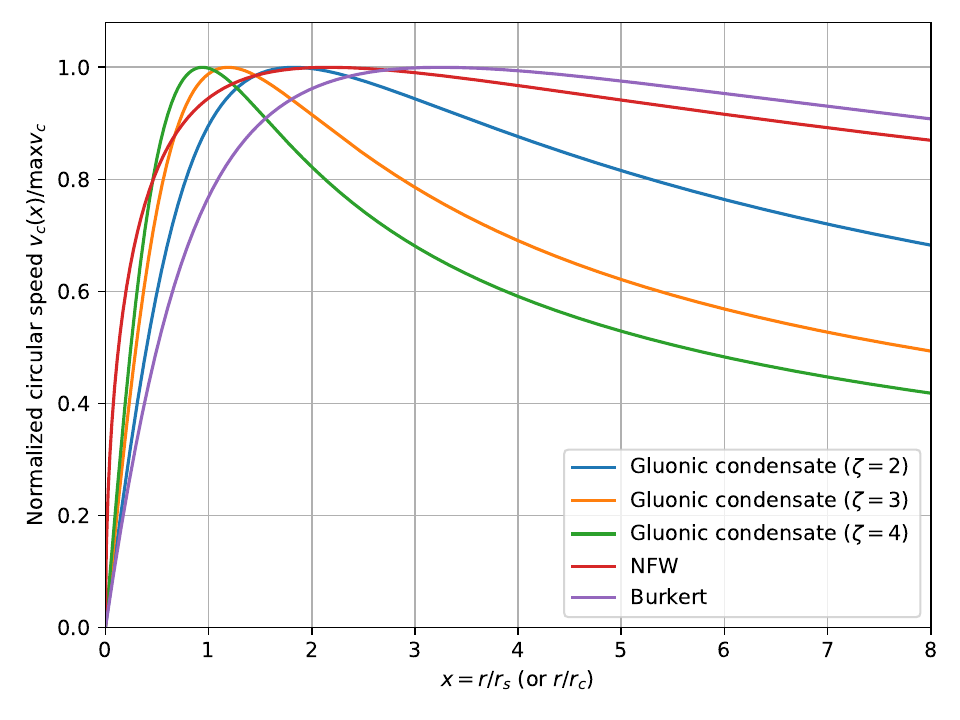}
    \caption{On the left, dimensionless rotation curves $v(x,\zeta)/\sqrt{4\pi G\rho^{}_{\texttt{c}} r_{\rm c}^2}$ as a function of $x=r/r_{\rm c}$ for $\zeta=2,\;5/2,\;3,\;4$. The inner linear rise reflects the constant-density core, while the outer Keplerian decline follows from the finiteness of the total halo mass. Larger values of $\zeta$ lead to lower and more centrally concentrated velocity profiles. On the right, shape comparison of circular-velocity profiles $v_{\texttt{c}}(r)$. The AdS-gluonic-condensate halo (shown for $\zeta=2,3,4$) is compared with the NFW and Burkert profiles. Each curve is normalized to its maximum value on the plotted interval in order to emphasize differences in shape independently of the amplitude set by $(\rho_s,r_s)$ or $(\rho_0,r_0)$.}
    \label{fig:rotation_curves1}
\end{figure*}

\begin{figure*}[t]
    \centering
    \includegraphics[width=0.55\linewidth]{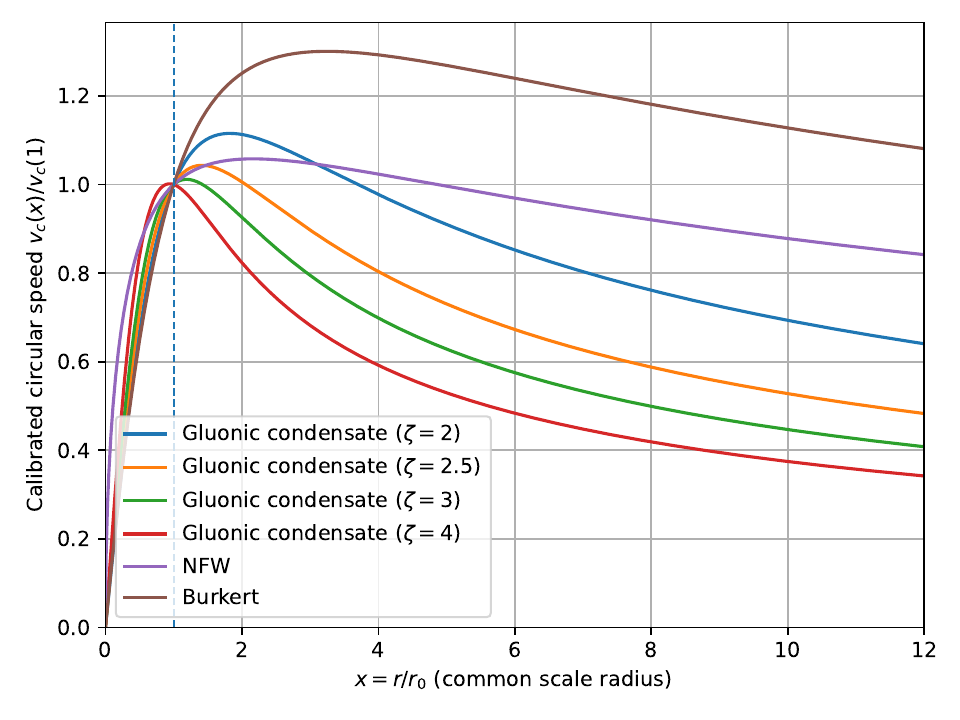}
    \caption{Calibrated comparison of circular-velocity profiles. All models are expressed in terms of the same dimensionless radius $x=r/r_0$ (common scale radius) and are normalized such that $v_{\texttt{c}}(1)$ is the same for all curves. The AdS-gluonic-condensate model is shown for $\zeta=2,\;5/2,\;3,\;4$ and compared with the NFW and Burkert profiles. The dashed vertical line marks $x=1$.}
    \label{fig:rotation_curves2}
\end{figure*}

\paragraph{Maximum of the rotation curve and comparisons with NFW \cite{Navarro1997} and Burkert \cite{Burkert1995} profiles.} As illustrated in Figs. \ref{fig:rotation_curves1} and \ref{fig:rotation_curves2} (see Tables \ref{tab:vc_formulas_summary1} and \ref{tab:vc_formulas_summary2} for the explicit formulas), the rotation velocity reaches a maximum at a dimensionless radius that we denote by $x^{}_{v;\max}$. From Eq. \eqref{eq:vdx}, the extremum condition $\mathrm{d}v/\mathrm{d}x=0$ immediately yields:
\begin{align} \label{eq:vmax_cond}
    \frac{F(x^{}_{v;\max},\zeta)}{x^{}_{v;\max}} = f(x^{}_{v;\max},\zeta) \,.
\end{align}
This relation admits a simple geometric interpretation. The quantity $F(x,\zeta)/x$ represents the slope of the straight line joining the origin to the point $(x, F(x,\zeta))$, while $f(x,\zeta)= F^\prime(x,\zeta)$ is the slope of the tangent to the curve $F(x,\zeta)$. Equation \eqref{eq:vmax_cond} therefore states that, at $x=x^{}_{v;\max}$, the tangent to the curve $F(x,\zeta)$ passes through the origin.

The velocity maximum occurs at a radius $x^{}_{v;\max}$ larger than the inflection point $x^{}_{M;\max}$, namely $x^{}_{v;\max}>x^{}_{M;\max}$. Since Eq. \eqref{eq:vmax_cond} admits no closed-form solution, this ordering is verified numerically, for example:
\begin{align}\label{eq:vmax}
    \zeta=2:&\quad x^{}_{M;\max}=1 \,, \quad x^{}_{v;\max}\approx 1.8253 \,, \nonumber\\
    \zeta=\frac52:&\quad x^{}_{M;\max}=\sqrt{\frac23}\,, \quad x^{}_{v;\max}=\sqrt2\,, \nonumber\\
    \zeta=3:&\quad x^{}_{M;\max}=\frac{1}{\sqrt2}\approx 0.7071 \,, \quad x^{}_{v;\max}\approx1.1892 \,, \nonumber\\
    \zeta=4:&\quad x^{}_{M;\max}=\frac{1}{\sqrt3} \approx 0.5773\,, \quad x^{}_{v;\max}\approx0.9403 \,.
\end{align}

Importantly, the dimensionless location $x^{}_{v;\max}$ depends only on the shape parameter $\zeta$ and is therefore universal within the model, while the corresponding physical radius is simply:
\begin{align}\label{eq:vmax'}
    r_{v;\max}=x^{}_{v;\max} \,r_{\rm c} \,.
\end{align}

\begin{table*}[t]
    \centering
    \renewcommand{\arraystretch}{2}
    \begin{tabular}{|l|l|l|}
    \hline
    \textbf{Model} & $\rho(x)$ & $M(<x)$  \\ \hline

    Gluonic condensate
    & $\rho^{}_{\texttt{c}}(1+x^2)^{-\zeta}$
    & $4\pi\rho^{}_{\texttt{c}} r_0^3\,F(x,\zeta)$
    \\ \hline

    NFW
    & $\dfrac{\rho_s}{x(1+x)^2}$
    & $4\pi\rho_s r_0^3\!\left(\ln(1+x)-\dfrac{x}{1+x}\right)$
    \\ \hline

    Burkert
    & $\dfrac{\rho_0}{(1+x)(1+x^2)}$
    & $\pi\rho_0 r_0^3\!\left(\ln(1+x^2)+2\ln(1+x)-2\arctan x\right)$
    \\ \hline
    \end{tabular}
    \caption{Summary of density profiles and enclosed masses. Here, $x=r/r_0$ denotes the dimensionless radius, where $r_0$ is the scale radius of each model ($r_0=r_{\rm c}$ for the gluonic condensate, $r_0=r_s$ for NFW, and $r_0=r_B$ for Burkert). For the AdS-gluonic-condensate model, $F(x,\zeta)=\int_0^x u^2(1+u^2)^{- \zeta}\,\mathrm{d}u$.}
    \label{tab:vc_formulas_summary1}
\end{table*}

\begin{table*}[t]
    \centering
    \renewcommand{\arraystretch}{2}
    \begin{tabular}{|l|l|l|}
    \hline
    \textbf{Model} & $\rho(x)$  & $v_{\texttt{c}}^2(x)=GM(<x)/x$ \\ \hline

    Gluonic condensate
    & $\rho^{}_{\texttt{c}}(1+x^2)^{-\zeta}$
    & $4\pi G\rho^{}_{\texttt{c}} r_0^2\,F(x,\zeta)/x$
    \\ \hline

    NFW
    & $\dfrac{\rho_s}{x(1+x)^2}$
    & $4\pi G\rho_s r_0^2\,\dfrac{\ln(1+x)-x/(1+x)}{x}$
    \\ \hline

    Burkert
    & $\dfrac{\rho_0}{(1+x)(1+x^2)}$
    & $\pi G\rho_0 r_0^2\,\dfrac{\ln(1+x^2)+2\ln(1+x)-2\arctan x}{x}$
    \\ \hline
    \end{tabular}
    \caption{Summary of density profiles and circular-speed laws in the same conditions as in Table \ref{tab:vc_formulas_summary1}.}
    \label{tab:vc_formulas_summary2}
\end{table*}

\subsection{Radial Accelerations} \label{subsec:radial_accelerations}

The radial gravitational acceleration generated by the DM halo in its ground state is defined as:
\begin{align} \label{eq:gdef_r}
    g(r,\zeta) = \frac{v^2(r,\zeta)}{r} = 4\pi G\rho^{}_{\texttt{c}} r_{\rm c}^3\, \frac{F\!\left(\frac{r}{r_{\rm c}},\zeta\right)}{r^2} \,.
\end{align}
In terms of the dimensionless radius $x=r/r_{\rm c}$, this reads:
\begin{align} \label{eq:gdef_x}
    g(x,\zeta) = 4\pi G\rho^{}_{\texttt{c}} r_{\rm c}\,\frac{F(x,\zeta)}{x^2} \,,
\end{align}
with radial derivative:
\begin{align} \label{eq:dgdx}
    \frac{\mathrm{d}g(x,\zeta)}{\mathrm{d}x} = 4\pi G\rho^{}_{\texttt{c}} r_{\rm c} \left( -2\,\frac{F(x,\zeta)}{x^3} + \frac{f(x,\zeta)}{x^2} \right) \,,
\end{align}
where, again, $f(x,\zeta)=x^2(1+x^2)^{-\zeta}$.

\paragraph{Central behavior.} As $x\ll 1$, using $F(x,\zeta)\sim x^3/3$, one finds:
\begin{align} \label{eq:gcore}
    g(x,\zeta) \sim \frac{4\pi G}{3}\,\rho^{}_{\texttt{c}}\, r_{\rm c}\,x \,,
\end{align}
corresponding to a linear growth of the acceleration. This behavior is independent of $\zeta$ and reflects once again the solid-body character of the constant-density core.

\paragraph{Asymptotic behavior.} For $x\gg1$ and $\zeta>\tfrac32$, the enclosed mass approaches a constant value, $F(x,\zeta)\longrightarrow F(\infty,\zeta)$, so that the acceleration becomes:
\begin{align} \label{eq:gasymp}
    g(x,\zeta) \sim \frac{G M_h(\zeta)}{r_{\rm c}^2\,x^2} \,,
\end{align}
which corresponds to the standard Newtonian inverse-square fall-off produced by a finite total halo mass.

\paragraph{Maximum of the radial acceleration.} The radial acceleration reaches a maximum at the radius $x^{}_{g;\max}$ determined by the condition:
\begin{align} \label{eq:gmax_cond}
    \frac{F(x^{}_{g;\max},\zeta)}{x^{}_{g;\max}} = \frac{1}{2}\, f(x^{}_{g;\max},\zeta) \,.
\end{align}
This relation fixes the position of the maximum of the radial acceleration. For the values of $\zeta$ considered in this work, one obtains:
\begin{align}\label{eq:gmax}
    \zeta=2:&\quad x^{}_{g;\max}\approx0.9682 \,, \nonumber\\
    \zeta=\frac52:&\quad x^{}_{g;\max}=1 \,, \nonumber\\
    \zeta=3:&\quad x^{}_{g;\max}\approx0.8260 \,, \nonumber\\
    \zeta=4:&\quad x^{}_{g;\max}\approx0.6766 \,,
\end{align}
which satisfy the ordering (see Eq. \eqref{eq:vmax}):
\begin{align}
    x^{}_{g;\max} < x^{}_{v;\max} \,,
\end{align}
showing that the maximum of the radial acceleration is reached at a smaller radius than the maximum of the circular velocity.

As in the case of the mass and velocity profiles, the dimensionless position $x^{}_{g;\max}$ depends only on the shape parameter $\zeta$ and is therefore universal, i.e., independent of the particular properties of a given galaxy. The corresponding physical radius is simply:
\begin{align}
    r_{\mathrm{g};\max}=x^{}_{g;\max}\,r_{\rm c} \,.
\end{align}

\begin{figure*}[t]
    \centering
    \includegraphics[width=0.5\linewidth]{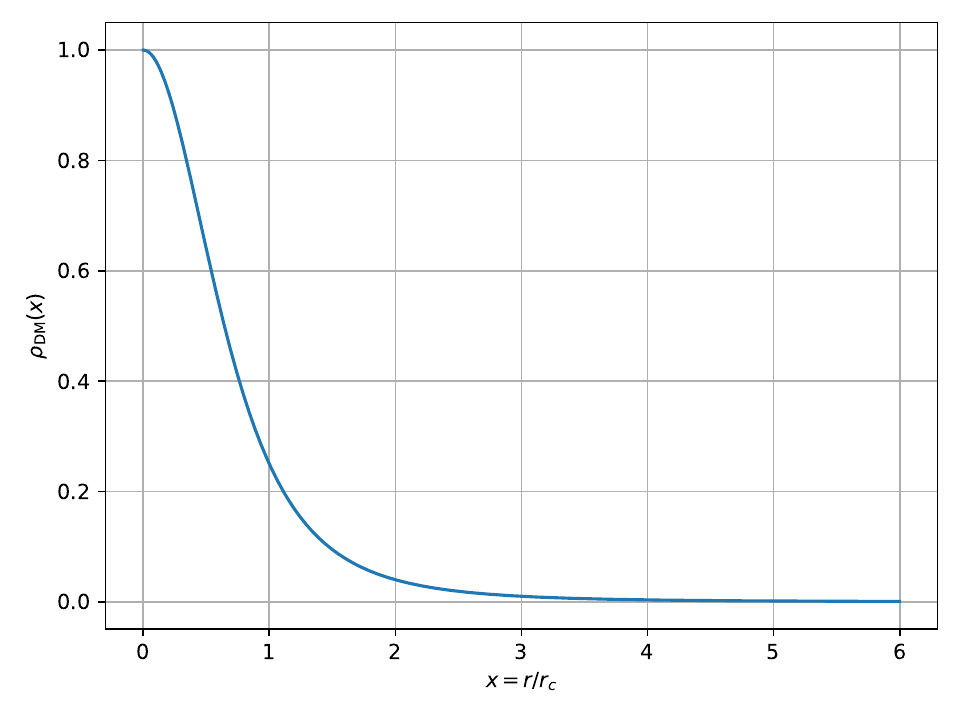}\hfill
    \includegraphics[width=0.5\linewidth]{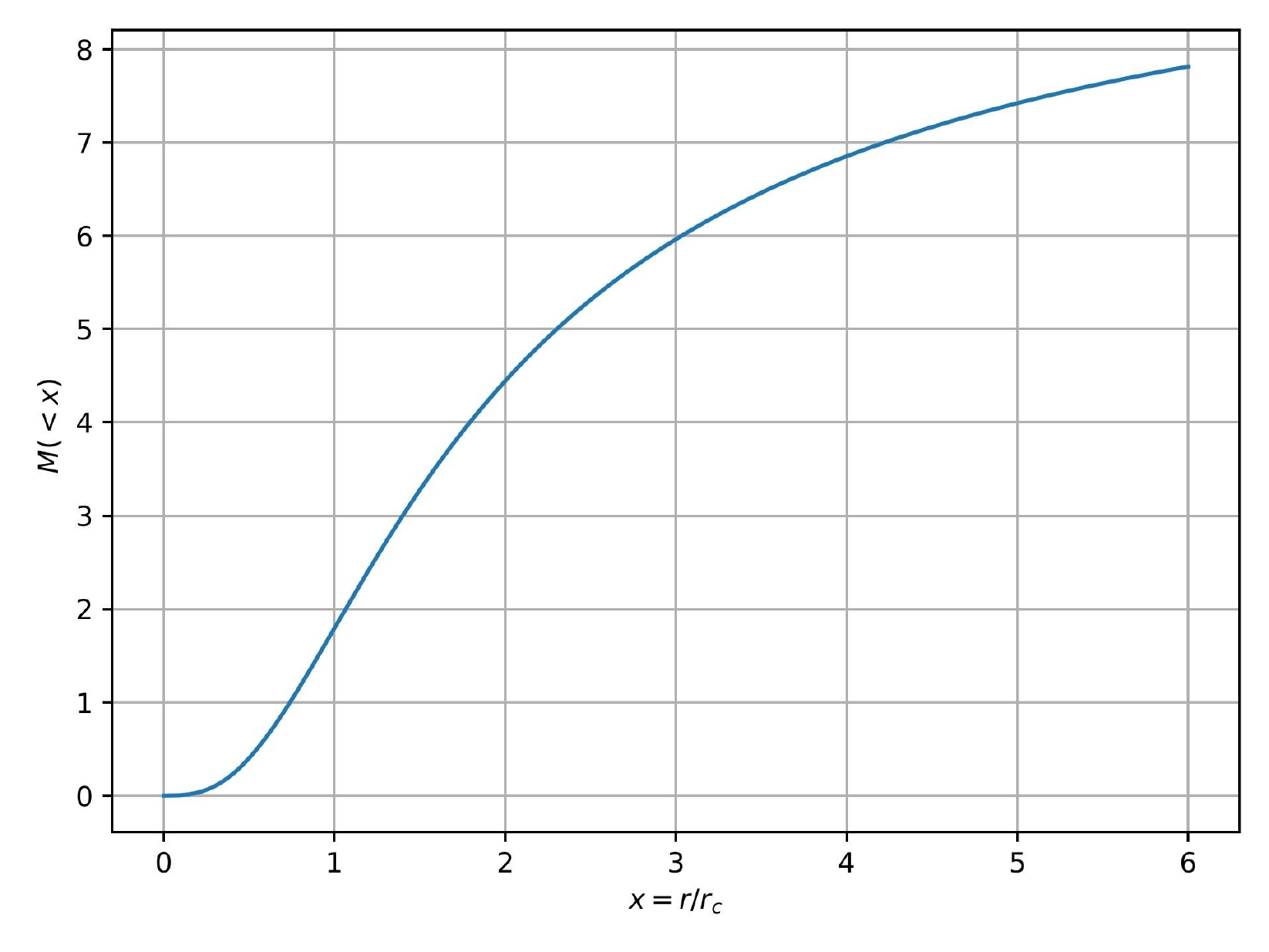}
    \caption{Illustrative density (left) and enclosed-mass (right) profiles for $\zeta=2$, in arbitrary units ($\rho^{}_{\texttt{c}}=r_{\rm c}=1$).}
    \label{fig:zeta2_density_mass}
\end{figure*}

\begin{figure*}[t]
    \centering
    \includegraphics[width=0.5\linewidth]{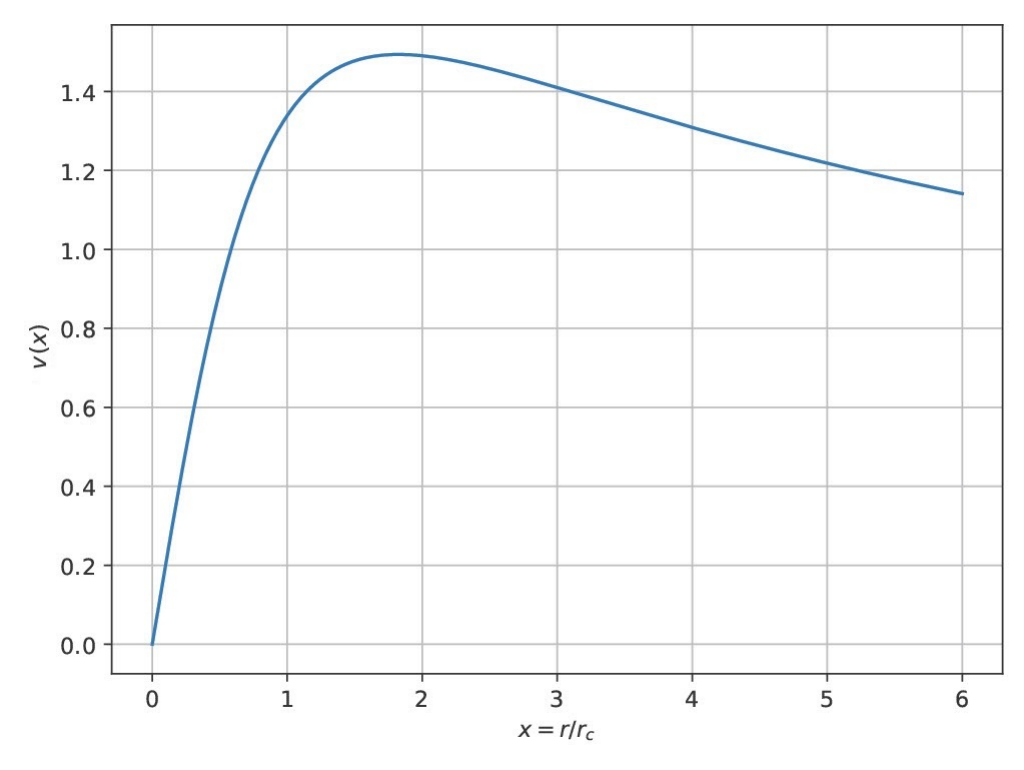}\hfill
    \includegraphics[width=0.5\linewidth]{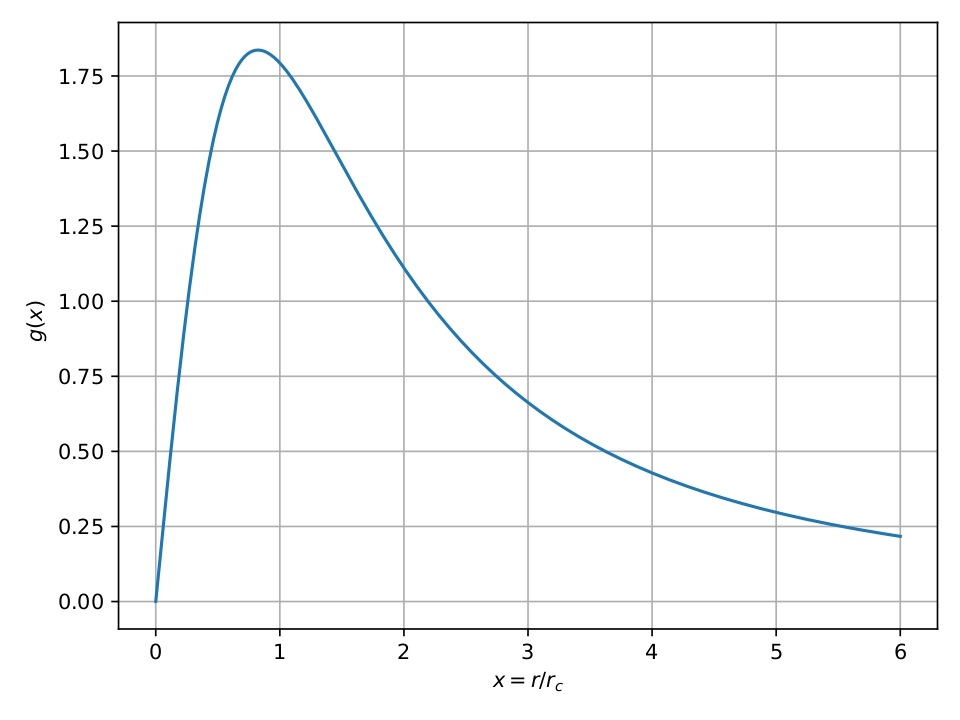}
    \caption{Illustrative circular-velocity (left) and radial-acceleration (right) profiles for $\zeta=2$, in arbitrary units ($\rho^{}_{\texttt{c}}=r_{\rm c}=1$).}
    \label{fig:zeta2_velocity_acceleration}
\end{figure*}

\subsection{On the Core Radius} \label{subsec:core_radius}

The previous analysis shows that the notion of a ``core radius'' is not unique and can be defined in several physically meaningful ways. Within the present framework, at least four natural definitions can be identified:
\begin{enumerate}
    \item{The value of $x$ beyond which the central approximation $M_{\mathrm{DM}}(<x,\zeta)\sim \frac{4}{3}\pi\rho^{}_{\texttt{c}} r_{\rm c}^3 x^3$ \eqref{eq:core} is no longer valid, signaling the breakdown of the constant-density regime.}

    \item{The value of $x=x^{}_{M;\max}$ \eqref{eq:xmax} corresponding to the inflection point of the enclosed-mass profile, where the curvature of $M_{\mathrm{DM}}(<x,\zeta)$ changes sign and the mass growth begins to slow.}

    \item{The value of $x=x^{}_{v;\max}$ \eqref{eq:vmax_cond} (see also \eqref{eq:vmax}) at which the circular velocity $v(x,\zeta)$ reaches its maximum.}

    \item{The value of $x=x^{}_{g;\max}$ \eqref{eq:gmax_cond} (see also \eqref{eq:gmax}) at which the radial acceleration $g(x,\zeta)$ attains its maximum.}
\end{enumerate}

These definitions are not identical, but they all select radii of the same order, thereby identifying a characteristic halo scale set by $r_{\rm c}$. This reflects the existence of an intermediate radial region separating the inner constant-density core from the outer domain in which the dynamics approaches the asymptotically Newtonian regime.

No specific baryonic response model is assumed in the present spherical analysis. Nevertheless, this intermediate radial regime is physically distinguished: it is where the ground-state profile departs from its constant-density core and where the enclosed-mass growth begins to slow. Small perturbations of the condensate can therefore have a comparatively large effect on the local density and acceleration profiles. This makes the region a natural domain in which baryonic perturbations, if included, would be expected to couple efficiently to the lowest non-spherical or radial excitations of the condensate.

\section{Baryonic Perturbations and Lowest Condensate Modes} \label{sec:baryonic_response}

In the absence of baryons, the dark condensate is described by the spherically symmetric ground state \eqref{ground}, whose density profile defines the characteristic scale $r_{\rm c}=\varkappa^{-1}$. Baryonic structures generate an additional weak gravitational potential which, in the present effective description, acts as a perturbation of this ground-state configuration. To leading order, the condensate response may therefore be represented by a truncated expansion over the lowest AdS modes:
\begin{align}
    \Psi_{\mathrm{DM}}(t,\mathbf{x}) \sim&\; a^{}_{000}\varphi^{}_{000} \nonumber\\
    &+ \sum_{m=-1}^{1} a^{}_{01m}\varphi^{}_{01m} + a^{}_{100}\varphi^{}_{100} + \cdots \,,
\end{align}
where the coefficients are determined by matrix elements of the baryonic perturbing potential. This expression should not be understood as a complete linear-response calculation, but as the minimal mode content expected to control the leading deformation of the condensate.

As anticipated in Sec. \ref{subsec:core_radius}, the relevant response is expected to be localized primarily in the transition region around $r\sim r_{\rm c}$. In the present section, this statement acquires a mode-theoretic interpretation; the same radial domain in which the ground-state profile becomes most susceptible to perturbations is also where the lowest non-spherical and radial excitations provide the leading response channels to baryonic gravity.

The first angular excitations are the triplet $(n,l)=(0,1)$ (see Eq. \eqref{01m}):
\begin{align}
    \varphi^{}_{01m}(r,\theta,\phi) \propto Y_{1m}(\theta,\phi)\, \frac{\varkappa r}{\left(1+(\varkappa r)^2\right)^{\tfrac{1}{2}(\zeta+1)}} \,.
\end{align}
They provide the leading non-spherical response channel. The factor $\varkappa r$ suppresses these modes at the origin and shifts their support toward $r\sim r_{\rm c}$, making them well suited to describe dipolar distortions induced by asymmetric baryonic structures. Physically, these modes encode the polarization of the condensate around baryonic concentrations, directional distortions aligned with the baryonic gravitational field, and possible coherent wakes associated with moving baryonic objects.

The first radial excitation, $(n,l)=(1,0)$, provides the leading spherically symmetric deformation: 
\begin{align}
    \varphi^{}_{100}(r) \propto \left((\varkappa r)^2-\frac{3}{2\zeta-1}\right) \frac{1}{\left(1+(\varkappa r)^2\right)^{\tfrac{1}{2}(\zeta+2)}} \,.
\end{align}
It has a radial node at:
\begin{align}
    r_{\rm node} = \varkappa^{-1}\sqrt{\frac{3}{2\zeta-1}} \,.
\end{align}
For the benchmark value $\zeta=2$, this gives $r_{\rm node}=r_{\rm c}$. A small admixture of this mode can therefore redistribute mass across the characteristic halo scale. At the level of the density, one may schematically write:
\begin{align}
    \rho^{}_{\mathrm{DM}}(r) \;\sim\; \mathcal{N} \left( |\varphi^{}_{000}(r)|^2 + \epsilon |\varphi^{}_{100}(r)|^2 \right) \,, \quad \epsilon\ll1 \,,
\end{align}
or, more generally, include the interference terms generated by the full superposition. Such radial admixtures can slightly modify the transition between the inner solid-body rise and the outer declining regime, while leaving the central constant-density core essentially unchanged.

These lowest modes, therefore, provide the natural channels through which baryons can perturb the condensate: the $(0,1)$ triplet governs the leading multipolar deformation, whereas the $(1,0)$ mode governs the leading breathing-type redistribution of mass. Their role in the present paper is structural. They show that the finite-mass ground-state halo is not a rigid background, but a coherent medium with a controlled low-energy susceptibility to baryonic perturbations. A quantitative derivation of the induced amplitudes $a^{}_{01m}$ and $a^{}_{100}$, and hence of the detailed baryon--DM acceleration relation, requires a dedicated linear-response calculation and is left for future work.

\section{RAR and Emergent Acceleration Scale} \label{sec:RAR}

The empirical RAR provides the main phenomenological motivation for comparing the intrinsic halo acceleration scale derived above with the observed galactic acceleration scale. SPARC analyses reveal a tight correlation between the observed acceleration $g^{}_{\mathrm{obs}}=v_{\mathrm{obs}}^2/r$ and the Newtonian acceleration generated by baryons, characterized by a scale:
\begin{align}
    g^{}_\dagger \simeq (1\text{--}2) \times 10^{-10} \,\mathrm{m\,s^{-2}} \,,
\end{align}
with small intrinsic scatter across a broad range of galaxies \cite{McGaugh2016, lelli2016SPARC, Lelli2017}. In the present work, however, we do not attempt to derive the full RAR. Rather, we use it only as an empirical benchmark for the acceleration scale generated by the gluonic condensate.

Within standard Newtonian gravity, we assume the usual decomposition:
\begin{align}
    g^{}_{\mathrm{obs}}(r) = g^{}_{\mathrm{bar}}(r) + g^{}_{\mathrm{DM}}(r)\,.
\end{align}
This framework is compatible with interpretations in which the RAR is viewed as an empirical manifestation of baryon-halo coupling within standard gravity, rather than necessarily requiring a modification of the gravitational law \cite{Navarro2017}.

The empirical relation may then be written in terms of the DM contribution as:
\begin{align}
    \frac{g^{}_{\mathrm{DM}}}{g^{}_\dagger} = \frac{g^{}_{\mathrm{bar}}}{g^{}_\dagger} \;\mu\!\left( \frac{g^{}_{\mathrm{bar}}}{g^{}_\dagger} \right) \,,
\end{align}
where $\mu$ denotes the corresponding empirical response function \cite{McGaugh2016,Lelli2017}. This form is useful here only because it emphasizes the existence of a characteristic acceleration scale controlling the transition between baryon-dominated and dark-matter-dominated regimes.

The question for the present framework is therefore whether a comparable scale emerges naturally from the condensate halo itself. To address this, we return to the acceleration profile derived above. As discussed previously, the radial acceleration produced by the ground-state halo can be written as (see Eqs. \eqref{eq:gdef_x}, \eqref{eq:Mx}, and \eqref{eq:Mreg})
\begin{align}\label{eq:gDM_general}
    g^{}_{\mathrm{DM}}(x,\zeta) \;\left(=g(x,\zeta)\right) &= 4\pi G\rho^{}_{\texttt{c}}r_{\rm c} \frac{F(x,\zeta)}{x^2} \nonumber\\
    &= \frac{G M_{\mathrm{DM}}(<x,\zeta)}{r_{\rm c}^2\, x^2} \nonumber\\
    &= g^{}_\star(\zeta)\, \frac{1}{x^2}\, \frac{F(x,\zeta)}{F(\infty,\zeta)} \,,
\end{align}
where (see Eq. \eqref{eq:Mh}):
\begin{align}\label{gStar}
    g^{}_\star(\zeta) = \frac{GM_h(\zeta)}{r_{\rm c}^2} &= 2\pi G\Sigma_0 \,B\!\left( \tfrac32, \zeta-\tfrac32 \right) \nonumber\\
    &= \pi^{3/2}G\Sigma_0 \,\frac{ \Gamma(\zeta-\frac32)} {\Gamma(\zeta)} \,,
\end{align}
with $\Sigma_0=\rho_c r_{\rm c}$. This expression is valid throughout the finite-mass halo sector $\zeta>3/2$.

The appearance of the combination $GM_h/r_{\rm c}^2$ therefore identifies a natural intrinsic acceleration scale associated with the finite halo mass and the collective correlation length. To estimate it phenomenologically, we use the observed approximate universality of the central DM surface density:
\begin{align}
    \Sigma_0 = \rho_c r_{\rm c} \simeq 141\,M_\odot\,\mathrm{pc}^{-2} \simeq 0.29\,\mathrm{kg\,m^{-2}} \,,
\end{align}
as inferred from cored-halo fits to galactic rotation curves \cite{KormendyFreeman2004,Donato2009,Gentile2009}. This density-length product is not a directly model-independent observable, since $r_{\rm c}$ is extracted from a chosen halo parametrization. Nevertheless, its approximate universality remains a robust phenomenological feature of cored halo descriptions, despite known departures in some systems \cite{Gopika2021}.

Substituting this phenomenological input into Eq. \eqref{gStar}, one finds that the dependence on the halo size cancels identically:
\begin{align}
    g^{}_\star(\zeta) = C(\zeta)\,G\Sigma_0 \,, \quad C(\zeta) = \pi^{3/2} \frac{\Gamma(\zeta-\frac32)}{\Gamma(\zeta)} \,.
\end{align}
Thus, within the finite-mass halo sector, an approximately universal density-length product is mapped into an intrinsic acceleration scale. The dimensionless coefficient $C(\zeta)$ is determined by the profile parameter $\zeta$, encoding the dependence of the intrinsic acceleration scale on the finite-mass condensate profile.

For the illustrative lowest-weight benchmark $\zeta=2$, one obtains:
\begin{align}
    C(2) = \pi^2 \,,
\end{align}
so that:
\begin{align}
    g^{}_\star(2) = \pi^2G\Sigma_0 \simeq 1.9\times10^{-10}\,\mathrm{m\,s^{-2}} \,.
\end{align}
This value lies within the empirical range:
\begin{align}
    g^{}_\dagger \simeq (1\text{--}2)\times10^{-10}\,\mathrm{m\,s^{-2}} \,,
\end{align}
inferred from SPARC analyses of the RAR \cite{McGaugh2016, Lelli2017}.

\begin{figure*}[t]
    \centering
    \includegraphics[width=1.0\linewidth]{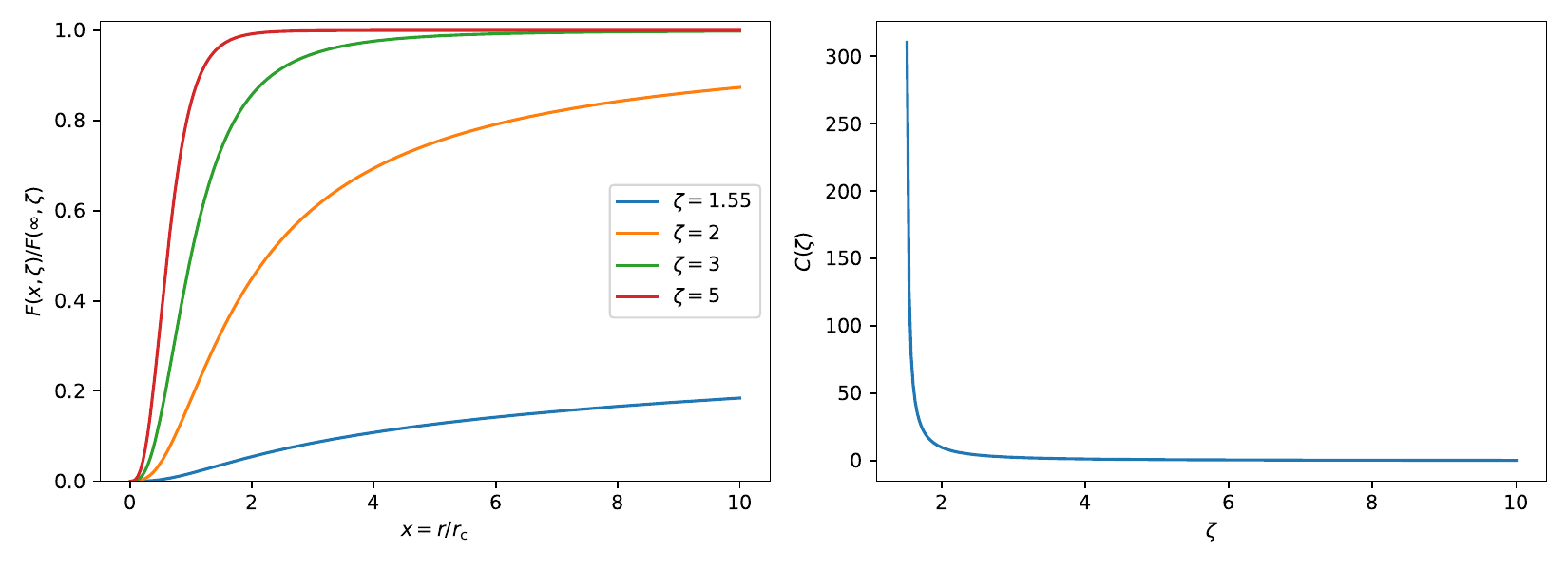}
    \caption{Left: Dimensionless enclosed-mass fraction $F(x,\zeta)/F(\infty,\zeta)$ as a function of $x=r/r_{\rm c}$ for representative values of the profile parameter $\zeta$. Profiles near the finite-mass threshold $\zeta\longrightarrow3/2^+$ saturate slowly, corresponding to extended and diffuse halo configurations, whereas larger values of $\zeta$ produce increasingly rapid saturation and more compact, centrally concentrated halos. Right: Dimensionless coefficient $C(\zeta)=\pi^{3/2}\Gamma(\zeta-\frac32)/\Gamma(\zeta)$ governing the intrinsic acceleration scale $g^{}_\star=C(\zeta)G\Sigma_0$. The benchmark value $C(2)=\pi^2$ gives $g^{}_\star(2)=\pi^2G\Sigma_0$, while the divergence as $\zeta\longrightarrow3/2^+$ reflects the approach to the boundary of the finite-mass sector.}
    \label{fig:two_panel_zeta_halo_structure}
\end{figure*}

It is worth noting that, apart from the near-critical region $\zeta\longrightarrow3/2^+$, where $C(\zeta)$ diverges, the coefficient $C(\zeta)$ varies smoothly and remains of the same order as its benchmark value $C(2)=\pi^2$. Thus, over a broad finite-mass range, the profile parameter primarily controls the spatial concentration of the halo, while the intrinsic acceleration scale remains parametrically comparable to $g^{}_\star(2)$. This behaviour is illustrated in Fig. \ref{fig:two_panel_zeta_halo_structure}.

It is also useful to express the halo acceleration in dimensionless form:
\begin{align}
    \frac{g^{}_{\mathrm{DM}}(x,\zeta)} {g^{}_\star(\zeta)} = \frac{1}{x^2}\, \frac{F(x,\zeta)}{F(\infty,\zeta)} \,.
\end{align}
This is the natural analogue, within the present halo construction, of the normalized quantity $g^{}_{\mathrm{DM}}/g^{}_\dagger$ used in empirical RAR studies. It makes explicit how the profile parameter controls the radial distribution of the dark acceleration. 

Near the boundary of the finite-mass sector, $\zeta=3/2+\epsilon$ with $\epsilon>0$, the total-mass normalization $F(\infty,\zeta)$ grows without bound, whereas $F(x,\zeta)$ remains finite at fixed $x$. Consequently:
\begin{align}
    \frac{g^{}_{\mathrm{DM}}(x,\zeta)}{g^{}_\star(\zeta)} \longrightarrow 0\,, \quad\mbox{for}\quad \zeta\longrightarrow 3/2^+ \,,
\end{align}
for any fixed radius. The normalized acceleration is therefore progressively suppressed at fixed radius, implying that an increasingly smaller fraction of the total halo mass is enclosed within a given radial scale.

The physical origin of this behavior becomes clearer from the asymptotic profile. At large radius:
\begin{align}
    \rho(r)\sim \rho_c\, x^{-2\zeta}\,, \quad x=\frac{r}{r_{\rm c}} \,,
\end{align}
so that the mass remaining outside a radius $x$ scales as:
\begin{align}
    M_h - M_{\rm DM}(<x) \propto \int_x^\infty y^{2-2\zeta}\,\mathrm{d}y = \frac{x^{3-2\zeta}}{2\zeta-3} \,, \quad \zeta>\frac32 \,.
\end{align}
For $\zeta=3/2+\epsilon$, this tail behaves as:
\begin{align}
    M_h - M_{\rm DM}(<x) \propto \frac{x^{-2\epsilon}}{2\epsilon} \,,
\end{align}
which decreases only slowly with radius. A substantial fraction of the total mass therefore remains distributed over large distances, supporting the interpretation that the near-critical regime $\zeta\simeq3/2$ corresponds to extended and diffuse halo configurations.

Conversely, for $\zeta\gg3/2$, the outer tail is rapidly suppressed:
\begin{align}
    M_h - M_{\rm DM}(<x) \propto \frac{x^{-(2\zeta-3)}}{2\zeta-3} \,.
\end{align}
Thus, $F(x,\zeta)$ approaches $F(\infty,\zeta)$ over a relatively short radial range. Equivalently, once $x$ lies beyond the region where the density is appreciable:
\begin{align}
    \frac{F(x,\zeta)}{F(\infty,\zeta)} \longrightarrow 1\,,
\end{align}
and hence:
\begin{align}
    \frac{g^{}_{\mathrm{DM}}(x,\zeta)}{g^{}_\star(\zeta)} \longrightarrow \frac{1}{x^2} \,.
\end{align}
The normalized acceleration then reaches the Newtonian finite-mass form over progressively shorter radial distances, indicating that an increasingly larger fraction of the total halo mass is enclosed within the inner halo. Large values of $\zeta$ therefore correspond to more centrally concentrated and compact halo configurations.

The result should be understood not as a derivation of the RAR itself, but rather as the emergence of an intrinsic halo acceleration scale of the same order as $g^{}_\dagger$. A first-principles derivation of the full relation between $g^{}_{\mathrm{obs}}$ and $g^{}_{\mathrm{bar}}$ would require a dedicated linear-response analysis of the condensate under baryonic perturbations.

\section{Comparison with Representative SPARC Rotation Curves} \label{sec:SPARC}

The preceding analysis has established the intrinsic properties of the finite-mass condensate halo, including its mass profile, rotation curve, acceleration scale, and dependence on the profile parameter $\zeta$. A natural next question is whether the resulting profile remains compatible with observed galactic rotation curves. As an initial phenomenological consistency check, we compare the present halo construction with representative systems from the SPARC database \cite{lelli2016SPARC}, which provides high-quality rotation curves together with photometric decompositions into gas, stellar disk, and bulge components.

Within standard Newtonian gravity, the circular velocity is decomposed as:
\begin{align}
    v_{\rm obs}^2(r) = v_{\rm gas}^2(r) + \Upsilon_d\, v_{\rm disk}^2(r) + \Upsilon_b\, v_{\rm bulge}^2(r) + v_{\rm DM}^2(r) \,,
\end{align}
where $\Upsilon_d$ and $\Upsilon_b$ denote the stellar mass-to-light ratios of the disk and bulge, respectively. The dark contribution is generated by the condensate halo.

For the present profile, Eq. \eqref{eq:vdef} gives:
\begin{align}
    v_{\rm DM}^2(r,\zeta) = 4\pi G \rho_c \,r_{\rm c}^2 \,\frac{F(r/r_{\rm c},\zeta)} {r/r_{\rm c}} \,.
\end{align}
If $\rho_c$ and $r_{\rm c}$ are treated as independent quantities, the halo contribution is specified by $(\rho_c,r_{\rm c},\zeta)$. In the phenomenologically constrained version emphasized in the present work, however, one adopts the approximate surface-density regularity discussed earlier $\Sigma_0 = \rho_c\, r_{\rm c}$, so that the dark component is effectively controlled by $(r_{\rm c},\zeta)$, with $\rho_c=\Sigma_0/r_{\rm c}$. In this form, $r_{\rm c}$ fixes the collective correlation length and characteristic transition scale, while $\zeta$ governs the redistribution of the halo mass between the inner and outer regions and controls the asymptotic approach toward finite-mass saturation. The finite-mass condition $\zeta>3/2$ remains essential; unlike phenomenological profiles requiring an external truncation scale, the total halo mass remains finite intrinsically.

\begin{figure*}[t]
    \centering
    \includegraphics[width=\textwidth]{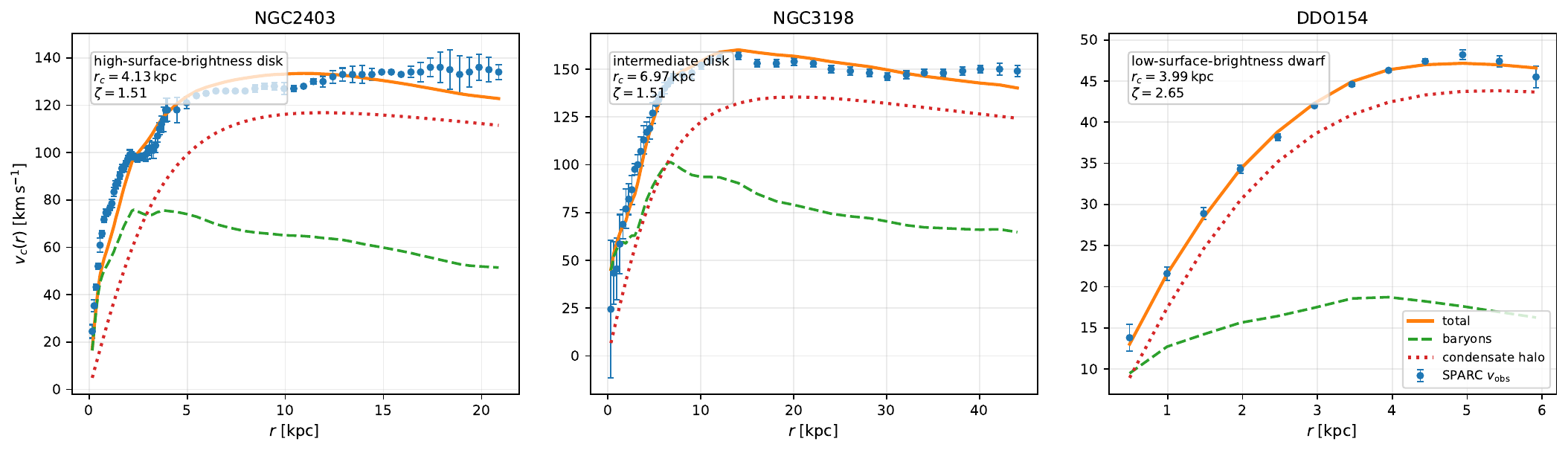}
    \caption{Illustrative comparison between the finite-mass gluonic-condensate halo profile and representative systems from the SPARC database \cite{lelli2016SPARC}. Points with error bars denote the observed circular velocities. Dashed curves show the baryonic contribution inferred from the SPARC decomposition into gas and stellar components, dotted curves represent the condensate-halo contribution, and solid curves show the total circular velocity. The examples span representative high-surface-brightness, intermediate, and low-surface-brightness systems. The comparison is intended as a representative consistency check rather than a precision statistical fit of the full SPARC sample.}
    \label{fig:SPARCfits}
\end{figure*}

Figure \ref{fig:SPARCfits} shows representative comparisons with SPARC galaxies spanning distinct baryonic morphologies. The baryonic contribution is taken directly from the SPARC decomposition into gas, disk, and bulge components, while the dark contribution is modeled by the finite-mass condensate profile derived above. The purpose of the comparison is not to establish a precision determination of model parameters, but rather to examine whether the finite-mass profile remains compatible with realistic galactic decompositions.

The representative examples indicate that the condensate profile is capable of reproducing the broad structure of observed rotation curves across systems with substantially different baryonic environments. In particular, more diffuse systems appear compatible with broader halo configurations, whereas more compact galaxies favor more centrally concentrated halo configurations. This trend is qualitatively consistent with the interpretation developed in previous sections; values approaching the finite-mass threshold $\zeta\to3/2^+$ generate increasingly extended halos, while larger values of $\zeta$ lead to more rapid mass saturation and stronger central localization.

A complete statistical analysis would require fits to the full SPARC sample, including observational uncertainties, stellar mass-to-light priors, parameter correlations, and comparison with standard halo parametrizations such as Burkert, NFW, and coreNFW profiles. Such an investigation lies beyond the scope of the present work and will be reported elsewhere. The present comparison should therefore be interpreted only as an indication that the finite-mass condensate profile is phenomenologically compatible with representative rotation-curve decompositions; it should not be viewed as a determination of model parameters or as a derivation of the RAR.

\section{Discussion and Outlook} \label{conclu}

The construction developed in this work should be understood as an effective infrared description of a coherent gluonic dark sector, not as a modification of the spacetime geometry probed by galactic tracers. The AdS structure is used internally; it organizes the long-wavelength color-singlet modes into a lowest-weight spectrum with a protected gap and an associated correlation length. This correlation length sets the halo scale entering the density profile derived above, but it does not imply that the physical galactic spacetime is AdS.

This separation of roles also justifies the Newtonian treatment used for rotation curves. Galactic dynamics is probed in the weak-field, slow-motion regime, where the standard relation between enclosed mass and circular velocity is the appropriate macroscopic closure. In the present framework, the AdS-based spectral structure fixes the internal organization of the dark component and hence its density profile, while baryonic tracers move in the ordinary Newtonian potential sourced by baryons and dark matter. Thus, no modification of gravity is assumed.

The main result is that the lowest scalar condensate mode yields a regular cored halo with finite total mass throughout the sector $\zeta>3/2$. This finiteness is essential; it permits the definition of an intrinsic halo acceleration scale without introducing an external truncation radius. When the profile is normalized by the observed approximate universality of the central dark-matter surface density, this scale becomes independent of the halo size. The conformal benchmark $\zeta=2$ gives a value lying in the empirical range of the RAR acceleration scale, while for a broad finite-mass range with $\zeta\gtrsim2$ the coefficient remains of the same order. Hence, the appearance of the observed acceleration scale is not tied to a sharp tuning of $\zeta$, but follows from the combination of finite total halo mass and the approximately universal density-length product $\Sigma_0$.

The near-critical regime $\zeta=3/2+\epsilon$ has a different interpretation. Although the total mass remains finite for every fixed $\epsilon>0$, it becomes increasingly dominated by the extended outer tail as $\epsilon\to0^+$. This regime therefore corresponds to diffuse, weakly concentrated halo configurations and marks the boundary of the finite-mass sector. By contrast, values $\zeta\gtrsim2$ describe more localized finite-mass halos, for which the acceleration scale remains stable while the profile parameter mainly controls the spatial concentration of the dark component.

The agreement with the RAR scale should therefore be viewed as the emergence of the correct characteristic acceleration, not as a first-principles derivation of the full RAR. A derivation of the relation between $g^{}_{\rm obs}$ and $g^{}_{\rm bar}$ requires the condensate response to baryonic perturbations. The mode analysis above identifies the natural starting point for such a calculation; the lowest dipolar modes provide the leading non-spherical response channels, while the first radial excitation controls the leading breathing-type redistribution of mass near the transition region $r\sim r_{\rm c}$. A quantitative linear-response calculation in this basis is therefore the next central step.

The comparison with representative SPARC rotation curves provides a complementary phenomenological check. It indicates that the same finite-mass profile can reproduce the broad structure of observed rotation curves when combined with realistic gas, disk, and bulge decompositions. This agreement should be interpreted conservatively; it demonstrates viability at the level of representative systems, not a global statistical validation of the model. A decisive observational assessment will require fitting the full SPARC sample, including uncertainties, stellar mass-to-light priors, and direct comparison with standard halo parametrizations.

The framework is also distinct from ultralight-boson or fuzzy-DM scenarios. In those models, macroscopic coherence is tied to a very small particle mass and a large de Broglie wavelength. Here, by contrast, coherence originates from the protected infrared spectral organization of a color-singlet gluonic sector, so that the halo scale is set by the lowest-weight correlation length rather than by an elementary-particle de Broglie scale.

Several issues remain open. The microscopic connection between the QCD trace-anomaly seed and the late-time lowest-weight infrared organization must be clarified. The formation, survival, and cosmological dilution of a color-singlet gluonic condensate from the confinement epoch to galactic times also require a dedicated treatment. Phenomenologically, the next step is a full statistical confrontation with high-quality rotation curves, beginning with the complete SPARC sample and extending to dwarf galaxies and systems with strongly non-spherical baryonic distributions. Such systems may be especially useful for testing whether dipolar or breathing-mode signatures of the condensate response provide distinctive observational discriminants.

In summary, the scenario proposed here connects the QCD trace anomaly, lowest-weight infrared spectral organization, finite-mass cored halos, and the emergence of a galactic acceleration scale of order $g^{}_\dagger$. In this picture, the cored density profile, finite total mass, approximate universality of the acceleration scale, and qualitative compatibility with representative rotation-curve decompositions are not independent assumptions, but different macroscopic manifestations of the same underlying infrared organization of the dark sector.

\section*{Acknowledgements}

Hamed Pejhan is supported by the National Science Fund, Ministry of Education and Science of Bulgaria, under contract KP-06-N92/2.


\end{document}